\documentclass[10pt,aps,pra,twocolumn,nofootinbib]{revtex4-1}   
\usepackage[dvips]{epsfig}
\usepackage{amsmath}    
\usepackage{amsfonts}  
\usepackage{amssymb}
\usepackage{graphicx}   
\usepackage{mathrsfs}
\usepackage{ulem}
\usepackage{mathtools}
\usepackage[usenames,dvipsnames]{color}

\usepackage[colorlinks]{hyperref}

\usepackage{colortbl}
\usepackage{multirow}

\usepackage{longtable}

\usepackage{braket}

\newcommand{\ignore}[1]{}

\usepackage{xcolor}

\AtBeginDocument{%
    \hypersetup{
      urlcolor     = blue, 
      filecolor   = red,
      linkcolor    = blue, 
      citecolor   = blue, 
    }
} 

\begin{document}

\title{Quantum Fisher information maximization in an unbalanced interferometer}

\author{Stefan~Ataman}
\affiliation{Extreme Light Infrastructure - Nuclear Physics (ELI-NP),\\ 
``Horia Hulubei'' National R\&D Institute for Physics and Nuclear Engineering (IFIN-HH),\\
30 Reactorului Street, 077125 Bucharest-M\u{a}gurele, Romania}
\email{stefan.ataman@eli-np.ro}

\date{\today}


\begin{abstract}
In this paper we provide the answer to the following question: given an arbitrary pure input state and a general, unbalanced, Mach-Zehnder interferometer, what transmission coefficient of the first beam splitter maximizes the quantum Fisher information (QFI)? We consider this question for both single- and two-parameter QFI, or, in other words, with or without having access to an external phase reference. We give analytical results for all involved scenarios. It turns out that, for a large class of input states, the balanced (50/50) scenario yields the optimal two-parameter QFI, however this is far from being a universal truth. When it comes to the single-parameter QFI, the balanced scenario is rarely the optimal one and an unbalanced interferometer can bring a significant advantage over the balanced case. We also state the condition imposed upon the input state so that no metrological advantage can be exploited via an external phase reference. Finally, we illustrate and discuss our assertions through a number of examples, including both Gaussian and non-Gaussian input states.

\end{abstract}

\maketitle

\section{Introduction}

Improving the interferometric phase sensitivity is both a classical and a quantum problem. The opportunity to surpass the classical shot-noise limit, and enter what is commonly called quantum or sub-shot-noise regime \cite{Cav81} is due to quantum metrology \cite{Giovannetti2006}. The boost of this field in recent years is correlated with the exponential growth of quantum technologies \cite{Acin2018}, however, it benefits from additional momentum from the gravitational wave astronomy community \cite{Tse19}, quantum-enhanced dark matter searches \cite{Backes2021} and QED (quantum electro-dynamics) vacuum phenomena \cite{Ata18b,Ahmadiniaz2020}.

The classical phase sensitivity limit $\Delta\varphi_{SQL}\sim1/\sqrt{\bar{N}}$ (also called standard quantum limit (SQL) where $\bar{N}$ denotes the average number of input photons) is a bound one gets with classical input states. However, by employing non-classical states of light \cite{Cav81}, the theoretically attainable limit shifts from SQL, to $\Delta\varphi_{HL}\sim1/{\bar{N}}$, also known as the Heisenberg limit (HL) \cite{Giovannetti2006}.


Among the available types of interferometric schemes we limit our discussion to the Mach-Zehnder interferometer (MZI) \cite{Loudon2003}, nonetheless the whole discussion can be adapted to other types of interferometers \cite{Dem15}. The balanced (50/50) MZI is usually discussed in the literature \cite{Lan13,Lan14,Pez08,Ata18,Ata19} and this scenario often yields simple expressions for the phase sensitivity. Besides, it is the optimal setup for a number of input states and detection schemes \cite{Ata18,Pez08,Lan13,Lan14,Gard2017}.


The question of optimal phase sensitivity for an interferometer arises, since one would like to optimize for all possible estimators and for all detection schemes. The elegant solution to this optimization problem is found by employing the quantum version of the classical Fisher information, namely the quantum Fisher information (QFI) \cite{Holevo1973,Braunstein1994,Paris2009}. Indeed, possesing the QFI, $\mathcal{F}$, allows one to employ the quantum Cram\'er-Rao bound (QCRB), $\Delta\varphi_{QCRB}=1/\sqrt{\mathcal{F}}$ \cite{Helstrom1967,Helstrom1968}. Thus, the phase sensitivity achievable by any realistic detection scheme $\Delta\varphi_{det}$ is bound to be $\Delta\varphi_{det}\geq\Delta\varphi_{QCRB}$.

Before moving on, one must mention other noteworthy approaches for optimal parameter estimation, including the Bayesian method \cite{Morelli2021,Dem15} or boson-sampling inspired strategies \cite{Valido2021}.


The realization that employing the QFI as defined above yields overly optimistic results \cite{Jar12} allowed to refine the analysis. Following Jarzyna and Demkowicz-Dobrza\'{n}ski \cite{Jar12}, we associate the asymmetric single-parameter QFI denoted by $\mathcal{F}^{(i)}$ to the scenario when a single phase shift $\varphi$ is applied inside the interferometer. For the scenario comprising two $\pm\varphi/2$ phase shifts (see Fig.~\ref{fig:MZI_2D_Fisher_info_sym_varphi_over2_plus_minus}) we associate the symmetric single-parameter QFI as $\mathcal{F}^{(ii)}$. Finally, by employing a two-parameter QFI we are able to denote the relevant difference-difference QFI denoted by $\mathcal{F}^{(2p)}$ \cite{Ata20}. Employing the two-parameter QFI guarantees that no external references are taken into account in the process of phase sensitivity evaluation.


As already discussed in the literature, the balanced case is not always optimal \cite{Pre19,Ata20} and unbalancing the interferometer can actually be beneficial \cite{Pre19,Ata20,Zhong2020}. When the interferometer is unbalanced, two supplementary parameters appear, namely the transmission coefficients of the first ($T$) and second ($T'$) beam splitters (BS).

As we will emphasize in this paper, optimizing the transmission coefficient of the first BS is enough to ensure the maximization of the QFI. This is due to the fact that the second BS has no effect whatsoever on the QFI calculation \cite{Zhong2020}. The statement remains true for both single- and two-parameter QFI scenarios, in other words, with or without an external phase reference. Thus, the question we set to answer in this work is what transmission coefficient $T$ optimizes each of the aforementioned QFIs, given a general input state, $\ket{\psi_{in}}$.


While optimum transmission coefficients for an unbalanced interferometer have been reported for some given constraints \cite{Zhong2020,Liu13} or for specific input states \cite{Ata20}, no universal solution to this problem was proposed, to the best of our knowledge. Moreover, in the previous studies \cite{Zhong2020,Liu13}, a specific QFI was considered only, namely the two-parameter QFI $\mathcal{F}^{(2p)}$.

In this paper we consider a general pure input state and an interferometric setup with/without external phase reference. We employ $\mathcal{F}^{(i)}$, $\mathcal{F}^{(ii)}$ and $\mathcal{F}^{(2p)}$ and each time we obtain an optimal beam splitter transmission coefficient. We show that the value of the optimal beam splitter transmission coefficient can always be analytically found.


Although not a central topic in this paper, the input phase matching conditions (PMC) \cite{Ata19,Ata20,Liu13} will be be discussed due to their connection to our optimization problem. We will show that different choices of QFI will point towards different optimal input PMCs.


After introducing the formal part, we go on to discuss these noteworthy scenarios. The first one deals with the conditions to be fulfilled by the input state in order to yield no quantum metrological advantage if an external phase reference is available. The second one involves an interformeter with one input in the vacuum state \cite{Takeoka2017}.


Finally, for a number of input states \cite{Campos1989,Par95,Anisimov2010,Spa15,Pre19,Ata19,Ata20,Birrittella2012,Birrittella2021}, we obtain the optimum transmission coefficient, $T_{opt}$, and thoroughly discuss the QFI performance for each scenario. Whenever possible, we compare our findings with previously reported ones in the literature.


Among the Gaussian input states, the squeezed-coherent plus squeezed-coherent input state was discussed previously for a balanced interferometer \cite{Spa15,Spa16,Ata19}. The same input state however employing the two-parameter QFI was considered at length in reference \cite{Ata19}, the discussion including the PMCs optimizing the two-parameter QFI as well as the performance of realistic detection schemes. Three input PMCs were singled out, each one maximizing $\mathcal{F}^{(2p)}$ in a certain regime. In references \cite{Spa15,Spa16} the authors employed the asymmetric single-parameter QFI and considered only the balanced case with an input PMC with all phases set to zero, except for one. As we will show in this paper, this PMC setting is not necessarily optimal, especially for the single-parameter QFI. To the best of our knowledge, there has been no discussion in the literature about the squeezed-coherent plus squeezed coherent input state applied to a non-balanced interferometer. We address this topic in this work and show that, similar to other Gaussian states \cite{Ata20}, using an unbalanced interferometer and having access to an external phase reference can bring a substantial increase in the QFI.

This paper is structured as follows. In Section \ref{sec:QO_description_MZI} we give the quantum optical description of our interferometer, introduce some notations and make some conventions. The Fisher matrix and the two-parameter QFI are both introduced in Section \ref{sec:QFI}. The two considered single-parameter QFIs are introduced in Section \ref{sec:single_param_QFI_F_i_and_F_ii}. The BS transmission coefficient maximizing each considered QFI is given in Section \ref{sec:optimal_T}, with all cases and sub-cases detailed. In Section \ref{sec:special_cases} we consider some noteworthy scenarios. Thoroughly discussed examples start in Section \ref{sec:examples}, where both Gaussian and non-Gaussian input states are evaluated. The paper ends with a short discussion in Section \ref{sec:discussion} followed by the conclusions from Section \ref{sec:conclusions}.

\section{The quantum optical description of an unbalanced MZI}
\label{sec:QO_description_MZI}
\begin{figure}
\includegraphics[scale=0.75]{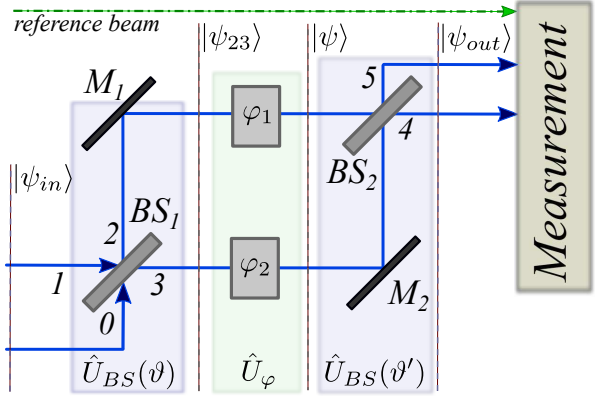}
\caption{A typical unbalanced Mach-Zehnder interferometric setup. In the general case we consider two independent phase shifts ($\varphi_1$ and $\varphi_2$) so that the effect of an eventual external phase reference can be revealed. When evaluating the QFI we call this setup ``closed'', as opposed to the one from Fig.~\ref{fig:MZI_Fisher_two_phases}.}
\label{fig:MZI_2D_closed_measure_at_output}
\end{figure}

In an interferometric setup one usually knows the input state vector $\ket{\psi_{in}}$ and wishes to determine the output one. If we consider more specifically a Mach-Zehnder interferometer (see Fig.~\ref{fig:MZI_2D_closed_measure_at_output}), the output state $\ket{\psi_{out}}$ can be formally written as
\begin{equation}
\label{eq:psi_out_U_BS_dagger_U_varphi_U_BS_psi_in}
\ket{\psi_{out}}=\hat{U}_{BS}\left(\vartheta'\right)\hat{U}_\varphi\hat{U}_{BS}\left(\vartheta\right)\ket{\psi_{in}}.
\end{equation}
We can model each beam splitter via the unitary operator
\begin{equation}
\label{eq:U_BS_exp_i_vartheta_Jx}
\hat{U}_{BS}\left(\varpi\right)=e^{i\varpi\hat{J}_x}
\end{equation}
where $\varpi\in\{\vartheta,\vartheta'\}$. If one wishes to connect the abstract angle $\vartheta$ to the more common beam splitter transmission coefficient, $T$, we can use the relation $\vartheta=2\arccos|T|$ and, similarly for $BS_2$, $\vartheta'=2\arccos|T'|$ \cite{Yurke1986,Campos1989}. $\hat{J}_x$ denotes the first Schwinger angular momentum operator \cite{Yurke1986,Campos1989},
\begin{equation}
\label{eq:QO_BS_and_MZI:Schwinger_J1_operator}
\hat{J}_x=\frac{\hat{a}_0^\dagger\hat{a}_1+\hat{a}_0\hat{a}_1^\dagger}{2}
\end{equation}
the other two being
\begin{equation}
\label{eq:QO_BS_and_MZI:Schwinger_J2_operator}
\hat{J}_y=\frac{\hat{a}_0^\dagger\hat{a}_1-\hat{a}_0\hat{a}_1^\dagger}{2i}
\end{equation}
and
\begin{equation}
\label{eq:QO_BS_and_MZI:Schwinger_J3_operator}
\hat{J}_z=\frac{\hat{a}_0^\dagger\hat{a}_0-\hat{a}_1^\dagger\hat{a}_1}{2}
\end{equation}
where $\hat{a}_l/\hat{a}_l^\dagger$ denote the usual annihilation/creation operators for the input modes $l=0,1$ \cite{GerryKnight}. We also introduce the input total photon number operator,
\begin{equation}
\label{eq:N_is_a_dagger_a_0_plus_1}
\hat{N}
=\hat{n}_0+\hat{n}_1
\end{equation}
where $\hat{n}_m=\hat{a}_m^\dagger\hat{a}_m$ denotes the usual number operator for a mode $m$. The three Schwinger angular momentum operators $\{\hat{J}_n|n\in\{x,y,z\}\}$ form a SU(2) Lie algebra (\emph{i. e.} $[\hat{J}_x,\hat{J}_y]=i\hat{J}_z$ etc.) and the Casimir element of the group is $\boldsymbol{\hat{J}}^2=\hat{N}/2(\hat{N}/2+1)$ \cite{Campos1989}. Please note that $\hat{N}$ commutes with all $\{\hat{J}_n|n\in\{x,y,z\}\}$ operators, a result that will be used in the following.

We allow two phase shifts in our MZI (see Fig.~\ref{fig:MZI_2D_closed_measure_at_output}), $\varphi_1$ ($\varphi_2$) in the upper (lower) arm of the interferometer. We also introduce the phase sum/difference \emph{i. e.} $\varphi_s=\varphi_1+\varphi_2$ and $\varphi_d=\varphi_1-\varphi_2$ variables. The effect of these phase shifts is modeled via the unitary operator
\begin{equation}
\label{eq:U_exp_minus_i_varphi_d_Jz}
\hat{U}_\varphi
=e^{-\varphi_1\hat{n}_2}e^{-i\varphi_2\hat{n}_3}
=e^{-\varphi_s\frac{\hat{N}}{2}}e^{-i\varphi_d\hat{J}_z}.
\end{equation}
Most authors prefer to add a fixed phase shift and thus model the effect of the MZI via ${\ket{\psi_{out}}=\hat{U}_{BS}^\dagger\left(\vartheta'\right)\hat{U}_\varphi\hat{U}_{BS}\left(\vartheta\right)\ket{\psi_{in}}}$. This is especially advantageous in the balanced case ($\vartheta=\vartheta'=\pi/2$) because the output state can be simply written as \cite{Liu13,Pez15,Yu2018}
\begin{equation}
\label{eq:psi_out_U_BS_dagger_U_varphi_U_BS_psi_in_BALANCED}
\ket{\psi_{out}}=e^{-\varphi_d\hat{J}_y}\ket{\psi_{in}}.
\end{equation}
For the general, non-balanced case, one can use the Euler-Rodrigues relations to simplify equation \eqref{eq:psi_out_U_BS_dagger_U_varphi_U_BS_psi_in} (see e. g. \cite{Campos1989} or Appendix A in reference \cite{Pez15}). As we will show in the following, for the sole purpose of QFI evaluation, these relations are not be needed.

\section{The Fisher matrix and the two-parameter quantum Fisher information}
\label{sec:QFI}
Before evaluating any QFI we remark that $\mathcal{F}$ being a measure of information is additive \cite{Paris2009}, thus repeating the same experiment $N$ times implies $\mathcal{F}\to N\mathcal{F}$. The implied QCRB, accordingly becomes $\Delta\varphi_{QCRB}=1/\sqrt{N\mathcal{F}}$. For simplicity, throughout this work, we consider $N=1$.

\subsection{Field operator transformations}
The annihilation field operators after the beam splitter $BS_1$ can be written as
\begin{equation}
\label{eq:a2_a3_U_BS_dagger_a1_a0_U_BS}
\left\{
\begin{array}{l}
\hat{a}_2=\hat{U}_{BS}^\dagger\left(\vartheta\right)\hat{a}_0\hat{U}_{BS}\left(\vartheta\right)=\cos\frac{\vartheta}{2}\hat{a}_0+i\sin\frac{\vartheta}{2}\hat{a}_1\\
\hat{a}_3=\hat{U}_{BS}^\dagger\left(\vartheta\right)\hat{a}_1\hat{U}_{BS}\left(\vartheta\right)
=i\sin\frac{\vartheta}{2}\hat{a}_0+\cos\frac{\vartheta}{2}\hat{a}_1\\
\end{array}
\right.
\end{equation}
relation that can be easily proven using the Baker-Hausdorf lemma \cite{GerryKnight}. If we use the parametrization $T=|T|=\cos\vartheta/2$ and $R=i|R|=i\sin\vartheta/2$ we get the usual field operator transformations for a symmetrical or ``thin-film'' beam splitter \cite{Loudon2003},
\begin{equation}
\label{eq:field_op_transf_MZI_a_hat}
\left\{
\begin{array}{l}
\hat{a}_2=T\hat{a}_0+R\hat{a}_1\\
\hat{a}_3=R\hat{a}_0+T\hat{a}_1
\end{array}
\right.
\end{equation}
where $T$ ($R$) denotes the transmission (reflection) of $BS_1$. For generic $T$ and $R$, energy conservation imposes the constraints $\vert{T}\vert^2+\vert{R}\vert^2=1$ and $TR^*+T^*R=0$ \cite{GerryKnight,Loudon2003}. Since the last relation implies $(T^*R)^2=-|TR|^2$, a sign convention has to be made (\emph{i. e.} $T^*R=\pm i|TR|$). From our parametrization we took the convention
\begin{equation}
\label{eq:iTstarR_convention}
iT^*R=-\vert{TR}\vert
\end{equation}
and it will be used throughout this work. In some optimizations we will use $|TR|$ as variable, we thus employ the replacement
\begin{equation}
\label{eq:T_and_R_as_fct_abs_TR}
\left\{
\begin{array}{ll}
\vert{T}\vert^2=\frac{1-\sqrt{1-4|TR|^2}}{2} & \textrm{ if } |T|^2\leq\frac{1}{2}\\
\vert{T}\vert^2=\frac{1+\sqrt{1-4|TR|^2}}{2} & \textrm{ if } |T|^2>\frac{1}{2}\\
\end{array}
\right.
\end{equation}
and the corresponding $\vert{R}\vert^2$ coefficients can be immediately deduced.

\begin{figure}
	\includegraphics[scale=0.75]{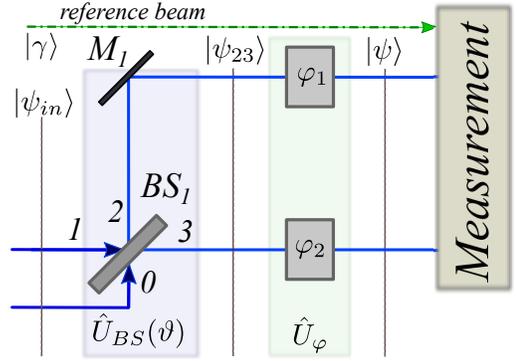}
	\caption{The ``open'' MZI setup considered when evaluating the QFI. The QFI evaluation is performed immediately after the two phase shifts.}
	\label{fig:MZI_Fisher_two_phases}
\end{figure}

\subsection{The Fisher matrix elements}
\label{subsec:the_Fisher_matrix}
We begin by discussing the two-parameter QFI. As pointed out previously in the literature \cite{Jar12,Lan13}, if one wishes to discharge any additional resources that potentially come from an external phase reference, a setup including two phase shifts must be discussed. The wavevector we consider, $\vert\psi\rangle=\hat{U}_\varphi\ket{\psi_{23}}$ (see Fig.~\ref{fig:MZI_Fisher_two_phases}), can be expressed in respect with the sum/difference phase shifts using equation \eqref{eq:U_exp_minus_i_varphi_d_Jz}, therefore
\begin{equation}
\label{eq:psi_phi_for_QFI}
\vert\psi\rangle
=e^{-i\hat{G}_d\varphi_d}e^{-i\hat{G}_s\varphi_s}\vert\psi_{23}\rangle
\end{equation}
where the generators are $\hat{G}_d=\hat{J}_z=(\hat{n}_2-\hat{n}_3)/{2}$ and $\hat{G}_s=\hat{N}/2=(\hat{n}_2+\hat{n}_3)/{2}$. We define the Fisher matrix elements \cite{Jar12,Lan13} (see also Appendix \ref{sec:app:QFI_Fisher_matrix}),
\begin{equation}
\label{eq:Fisher_matrix_elements}
\mathcal{F}_{ij}=4\Re\{\langle\partial_i\psi\vert\partial_j\psi\rangle-\langle\partial_i\psi\vert\psi\rangle
 \langle\psi\vert\partial_j\psi\rangle\}
\end{equation}
where $i, j\in \{s,d\}$, $\Re$ denotes the real part and we denote $\vert\partial_j\psi\rangle=\partial\vert\psi\rangle/\partial_{\varphi_j}$. The choice of the above definition of the QFI (\emph{i. e.} excluding $BS_2$) is justified in Appendix \ref{sec:app:open_vs_closed_MZI}, where it is shown that the effect of $BS_2$ on the QFI is null.
 
The first Fisher matrix element we consider is the so-called ``sum-sum'' coefficient and recalling that $[{\hat{J}_x},{\hat{N}}]=0$ we immediately have  (see also Appendix \ref{sec:app:F_ss_calculations}),
\begin{equation}
\label{eq:F_ss_DEF}
\mathcal{F}_{ss}=\braket{\psi_{in}|\hat{N}^2|\psi_{in}}-\braket{\psi_{in}|\hat{N}|\psi_{in}}^2=\Delta^2\hat{N}
\end{equation}
and by employing the appropriate shorthand notations from Appendix \ref{sec:app:Shorthand notations}, it reads
\begin{equation}
\label{eq:F_ss_shorthand_notations}
\mathcal{F}_{ss}=V_{+}+V_{cov}.
\end{equation}
The ``difference-difference'' Fisher matrix coefficient $\mathcal{F}_{dd}$ is found to be (see Appendix \ref{sec:app:F_dd_calculations})
\begin{equation}
\label{eq:Fdd_cos2_vartheta_Variance_Jz_sin2_vartheta_Variance_Jy}
\mathcal{F}_{dd}=4\left(
\cos^2\vartheta\Delta^2\hat{J}_z
+\sin^2\vartheta\Delta^2\hat{J}_y
-\sin2\vartheta\widehat{\text{Cov}}(\hat{J}_z,\hat{J}_y)
\right)
\end{equation}
where the symmetrized covariance is defined by
\begin{equation}
\label{eq:SymmetrizedCovariance_DEF}
\widehat{\text{Cov}}({\hat{J}_z},\hat{J}_y)=\frac{\braket{\hat{J}_z\hat{J}_y}+\braket{\hat{J}_y\hat{J}_z}}{2}-\braket{\hat{J}_y}\braket{\hat{J}_z}.
\end{equation}
Employing the appropriate shorthand notations from Appendix \ref{sec:app:Shorthand notations}, we are taken to
\begin{eqnarray}
\label{eq:F_dd_shorthand_notations_bis}
\mathcal{F}_{dd}=
V_+-V_{cov}
+\vert{TR}\vert^2\left(A-4(V_{+}-{V_{cov}})\right)
\nonumber\\
-2|TR|\left(\vert{T}\vert^2-\vert{R}\vert^2\right){S_{+}}.
\end{eqnarray}
The ``sum-difference'' Fisher matrix element $\mathcal{F}_{sd}$ is found to be
\begin{equation}
\label{eq:F_sd_Schwinger_FINAL}
\mathcal{F}_{sd}
=2\cos\vartheta\text{Cov}\left(\hat{N},\hat{J}_z\right)
-2\sin\vartheta\text{Cov}\left(\hat{N},\hat{J}_y\right)
\end{equation}
where the covariance of two operators $\hat{A}$ and $\hat{B}$ is defined by
\begin{equation}
\label{eq:covariance_DEF}
\textrm{Cov}(\hat{A},\hat{B})=\langle{\hat{A}}{\hat{B}}\rangle
-\langle{\hat{A}}\rangle\langle{\hat{B}}\rangle
\end{equation}
and details on the calculation of $\mathcal{F}_{sd}$ are given in Appendix \ref{sec:app:F_sd_calculations}. Using again the shorthand notations from Appendix \ref{sec:app:Shorthand notations}, we can write equation \eqref{eq:F_sd_Schwinger_FINAL} as
\begin{equation}
\label{eq:F_sd_shorthand_notations}
\mathcal{F}_{sd}
=\left(\vert{T}\vert^2-\vert{R}\vert^2\right)V_{-}
-\vert{TR}\vert\left(P+S_{-}\right).
\end{equation}
The fourth matrix element, $\mathcal{F}_{ds}$, is not needed, since $\mathcal{F}_{ds}=\mathcal{F}_{sd}$.

\subsection{The two-parameter difference-difference QFI}
\label{subsec:Fisher_2p}
We introduce now the quantum Fisher information relevant for a phase difference detection sensitivity \cite{Jar12,Lan13,Ata19,Ata20} (see also Appendix \ref{sec:app:QFI_Fisher_matrix}),
\begin{equation}
\label{eq:Fisher_information_F_2p_DEFINITION}
\mathcal{F}^{(2p)}
=\mathcal{F}_{dd}-\frac{\mathcal{F}_{sd}^2}{\mathcal{F}_{ss}}
\end{equation}
and this QFI implies the difference-difference QCRB,
\begin{equation}
\Delta\varphi_{QCRB}^{(2p)}=\frac{1}{\sqrt{\mathcal{F}^{(2p)}}}.
\end{equation}
We recall that this is the ``true'' interferometric phase sensitivity for a MZI with a detection scheme not having access to an external phase reference \cite{Jar12,Ata20}. When considering balanced interferometers with a given input state some authors take $\mathcal{F}_{sd}=0$, thus $\mathcal{F}^{(2p)}=\mathcal{F}_{dd}$ \cite{Lan13,Lan14}. Since our focus in on non-balanced scenarios, we will use the complete expression \eqref{eq:Fisher_information_F_2p_DEFINITION} throughout this work.

\section{The QFI for the single parameter cases}
\label{sec:single_param_QFI_F_i_and_F_ii}

\subsection{The asymmetric single parameter QFI, $\mathcal{F}^{(i)}$}
\label{subsec:F_i}
In this scenario we assume a single phase shift, $\varphi_2=\varphi$ and consequently $\varphi_1=0$ (see Fig.~\ref{fig:MZI_Fisher_two_phases}). We have the phase shift generator $\hat{G}=\hat{n}_3$ and thus the implied QFI can be simply expressed as \cite{Paris2009},
\begin{equation}
\label{eq:F_i_is_four_times_variance_n3}
\mathcal{F}^{(i)}=4\Delta^2{\hat{n}_3}.
\end{equation}
The QFI $\mathcal{F}^{(i)}$ implies the QCRB
\begin{equation}
\label{eq:Delta_varphi_QCRB_i_DEFINTION}
\Delta\varphi^{(i)}_{QCRB}=\frac{1}{\sqrt{\mathcal{F}^{(i)}}}
\end{equation}
and this scenario corresponds to the phase sensitivity when an external phase reference is available \cite{Jar12,Ata20}.

The calculation can be done by employing the field operator transformations \eqref{eq:a2_a3_U_BS_dagger_a1_a0_U_BS} and the result is given in Appendix \ref{sec:app:F_i_calculations}. An alternative method to calculate $\mathcal{F}^{(i)}$ is to take advantage of the relation connecting the assymetric single-parameter QFI $\mathcal{F}^{(i)}$ to the Fisher matrix coefficients \cite{Ata20},
\begin{equation}
\label{eq:F_i_is_Fss_plus_Fdd-2Fds}
\mathcal{F}^{(i)}=\mathcal{F}_{ss}+\mathcal{F}_{dd}-2\mathcal{F}_{sd}.
\end{equation}
Since not all authors consider the phase shift in the lower arm of the interferometer, we briefly consider the other possible convention for the asymmetric single-parameter QFI in Appendix \ref{sec:app:F_i_calculations_n2}.

\begin{figure}
\includegraphics[scale=0.75]{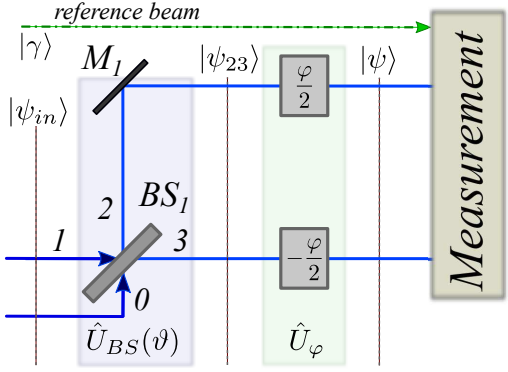}
\caption{Mach-Zehnder interferometric setup with symmetrical $\pm\varphi/2$ phase shifts. The QFI evaluating the performance of this setup is $\mathcal{F}^{(ii)}$.}
\label{fig:MZI_2D_Fisher_info_sym_varphi_over2_plus_minus}
\end{figure}

\subsection{The symmetric single parameter QFI $\mathcal{F}^{(ii)}$}
\label{subsec:F_ii}
If we assume $\varphi_1=\varphi/2$ and $\varphi_2=-\varphi/2$, we have the experimental setup depicted in Fig.~\ref{fig:MZI_2D_Fisher_info_sym_varphi_over2_plus_minus}. We have now\footnote{If we take the opposite convention \emph{i. e.} $\vert\psi\rangle=e^{i\frac{\varphi}{2}\left(\hat{n}_2-\hat{n}_3\right)}\vert\psi_{23}\rangle$ the QFI from equation \eqref{eq:Fii_variance_n3_minus_n2} remains obviously unchanged.} $\vert\psi\rangle=e^{-i\frac{\varphi}{2}\left(\hat{n}_2-\hat{n}_3\right)}\vert\psi_{23}\rangle$, the QFI is thus given by
\begin{equation}
\label{eq:Fii_variance_n3_minus_n2}
\mathcal{F}^{(ii)}=\Delta^2({\hat{n}_2}-{\hat{n}_3})
=\braket{({\hat{n}_2}-{\hat{n}_3})^2}-\braket{{\hat{n}_2}-{\hat{n}_3}}^2
\end{equation}
and this is actually the Fisher matrix element $\mathcal{F}_{dd}$, already computed in equation \eqref{eq:Fdd_cos2_vartheta_Variance_Jz_sin2_vartheta_Variance_Jy}. The achievable phase sensitivity in this scenario is lower bounded by the QCRB
\begin{equation}
\Delta\varphi^{(ii)}_{QCRB}=\frac{1}{\sqrt{\mathcal{F}^{(ii)}}}.
\end{equation}

\section{Optimum transmission coefficient}
\label{sec:optimal_T}
We are ready now to deduce the optimal transmission coefficient ($T_{opt}$) of the first beam splitter in the sense of maximizing each considered QFI. In order to lay bare the $T$-dependence of each QFI we introduce some coefficients that will allow an extremely compact writing of each expression.

\subsection{The two-parameter difference-difference QFI}
\label{subsec:Fisher_2p_optimal_T}
We start our discussion with the two-parameter QFI, $\mathcal{F}^{(2p)}$. From definition \eqref{eq:Fisher_information_F_2p_DEFINITION} and the using the appropriate expression for each Fisher matrix element we arrive at
\begin{equation}
\label{eq:F_2p_shorthand}
\mathcal{F}^{(2p)}=C_0+C_1\vert{TR}\vert^2+C_2\vert{TR}\vert\left(|T|^2-|R|^2\right)
\end{equation}
where the $C$-coefficients are given in Appendix \ref{sec:app:F_2p_calculations}.

We can discuss now the maximization of $\mathcal{F}^{(2p)}$ as a function of $|T|$. For simplicity, we consider $T$ real in the remainder of this section. 

\begin{table}
\centering
\renewcommand{\arraystretch}{1.3}
\begin{tabular}{c||c|c|c|c|}
\arrayrulecolor{black}
\cline{2-5}
 & \multicolumn{4}{c|}{constraints obeyed by the $C$-coefficients} \\
\cline{2-5}
\cline{2-5}
 & $C_1=0$ & $C_1\neq0$ & \multicolumn{2}{c|}{$C_2=0$} \\
\cline{4-5}
 & $C_2=0$ & $C_2\neq0$ &  $C_1>0$ & $C_1<0$ \\
\arrayrulecolor{black}
\hline
\multicolumn{1}{|c||}{$T^{(2p)}_{opt}$} & irrelevant & $\sqrt{\frac{1+\text{sgn}(C_2)\sqrt{\frac{1}{2}-\frac{\text{sgn}(C_1)|C_1|}{2\sqrt{C_1^2+4C_2^2}}}}{2}
}$ & $\frac{1}{\sqrt{2}}$  & $0/1$ \\
\hline
\multicolumn{1}{|c||}{$\mathcal{F}^{(2p)}_{max}$} & $C_0$ & $C_0+\frac{C_1}{8}
+\frac{\sqrt{C_1^2+4C_2^2}}{8}$ &  $C_0+\frac{C_1}{4}$   & $C_0$ \\
\hline
\end{tabular}
\caption{\label{tab:T_2p_and_F_2p}Optimal transmission coefficient $T^{(2p)}_{opt}$ and the corresponding maximum two-parameter QFI $\mathcal{F}^{(2p)}_{max}$ in all discussed scenarios.}
\end{table}


\noindent A) The case $C_1=C_2=0$ implies a constant QFI \emph{i. e.} $\mathcal{F}^{(2p)}=C_0$ and the value of $T$ becomes irrelevant. 

\noindent B) The simple case when $C_1\neq0$ but $C_2=0$ implies two scenarios:
\begin{enumerate}
	\item[i)] if $C_1>0$, then $\mathcal{F}^{(2p)}$ is maximized in the balanced case, \emph{i. e.} 
\begin{equation}
{T}_{opt}^{(2p)}=\frac{1}{\sqrt{2}}
\end{equation}	
This happens for a number of input states, and it is the most commonly discussed scenario in the literature \cite{Pez07,Lan13,Lan14,Pez15}.
	\item[ii)] if $C_1<0$, $\mathcal{F}^{(2p)}$ is maximized in the degenerate case (i.e. $T=0/1$). While not very common, this scenario can happen even for the coherent plus squeezed vacuum \cite{Pre19} and squeezed-coherent plus squeezed vacuum \cite{Ata20} inputs, given some input parameter choices.
\end{enumerate}
\noindent C) In the most general case when $C_1\neq0$ and $C_2\neq0$, we arrive at the optimal transmission coefficient (see Appendix \ref{sec:app:F_2p_calculations}),
\begin{equation}
\label{eq:T_opt_squared_F_2p}
{T}_{opt}^{(2p)}
=\sqrt{\frac{1+\text{sgn}(C_2)\sqrt{\frac{1}{2}-\frac{\text{sgn}(C_1)|C_1|}{2\sqrt{C_1^2+4C_2^2}}}}{2}
}
\end{equation}
yielding the maximum two-parameter QFI
\begin{equation}
\label{eq:F_2p_shorthand_MAX}
\mathcal{F}^{(2p)}_{max}=C_0+\frac{C_1}{8}
+\frac{\sqrt{C_1^2+4C_2^2}}{8}
\end{equation}
where $\text{sgn}$ denotes the signum function, \emph{i. e.} ${\text{sgn}:\mathbb{R}\to\{-1,0,1\}}$,  $\text{sgn}(x)=-1$ if $x<0$, $\text{sgn}(x)=+1$ if $x>0$ and $\text{sgn}(x)=0$ if $x=0$.

All results from this section are summarized in Table~\ref{tab:T_2p_and_F_2p}.

\begin{table*}
\centering
\renewcommand{\arraystretch}{1.3}
\begin{tabular}{c||c|c|c|c|c|c|c|c|c|}
\arrayrulecolor{black}
\cline{2-10}
 & \multicolumn{9}{c|}{constraints obeyed by the $C'$-coefficients} \\
\cline{2-10}
\cline{2-10}
& \multicolumn{2}{c|}{$C'_2=C'_3=C'_4=0$}  & \multicolumn{5}{c|}{$C'_2=C'_4=0$, $C'_3\neq0$} & \multicolumn{2}{c|}{$C'_1=C'_2=0$} \\
\cline{2-10} 
 & $C'_1<0$ & $C'_1>0$ & eq.~\eqref{eq:T_i_opt_existence_cond_C2prime_C4prime_ZERO} & eq.~\eqref{eq:T_i_opt_is_0_COND_C2prime_C4prime_ZERO} & eq.~\eqref{eq:T_i_opt_is_1_COND_C2prime_C4prime_ZERO} & eq.~\eqref{eq:T_i_opt_is_0or1_COND_C2prime_C4prime_ZERO},$C'_3<0$ & eq.~\eqref{eq:T_i_opt_is_0or1_COND_C2prime_C4prime_ZERO},$C'_3>0$ &  $C'_4>0$ & $C'_4<0$ \\
\arrayrulecolor{black}
\hline
\multicolumn{1}{|c||}{$T^{(i)}_{opt}$}  & $0/1$  & $\frac{1}{2}$ & $\sqrt{\frac{1}{2}+\frac{C'_3}{C'_1}}$ & 0 & 1 & 0 & 1 & $\sqrt{\frac{1}{2}+\frac{|C'_3|\text{sgn}(C'_3)}{\sqrt{4(C'_3)^2+(C'_4)^2}}}$   & $0/1$ \\
\hline
\multicolumn{1}{|c||}{$\mathcal{F}^{(i)}_{max}$} & $C'_0$ & $C'_0+\frac{C'_1}{4}$ & $C'_0+C'_1\left(\frac{1}{4}
+\frac{{C'_3}^2}{{C'_1}^2}
\right)$ & $C'_0-C'_3$ & $C'_0+C'_3$ & $C'_0-C'_3$ & $C'_0+C'_3$ &  $C'_0
+\frac{\sqrt{4(C_3')^2+(C_4')^2}}{2}$   & $C'_0\pm C'_3$ \\
\hline
\end{tabular}
\caption{\label{tab:T_i_and_F_i}Optimal transmission coefficient and the corresponding maximum asymmetric single-parameter QFI  in the discussed scenarios.}
\end{table*}

\subsection{The asymmetric single parameter QFI, $\mathcal{F}^{(i)}$}
\label{subsec:F_i_optimal_T}
The single-parameter QFI \eqref{eq:F_i_is_four_times_variance_n3} can be written as
\begin{eqnarray}
\label{eq:F_i_in_shorthand_C_prime}
\mathcal{F}^{(i)}=C'_0+\vert{TR}\vert^2 C'_1+|TR|(|T|^2-|R|^2)C'_2
\nonumber\\
+(|T|^2-|R|^2)C_3'+\vert{TR}\vert C_4'
\end{eqnarray}
where the $C'$-coefficients are given in Appendix \ref{sec:app:calc_T_i_opt_for_Fi}.

We start the optimality discussion with a less general case, however comprising a large number of input states\footnote{Among them, we mention the coherent plus squeezed vacuum \cite{Ata20,Pre19}, squeezed-coherent plus squeezed vacuum \cite{Ata20} (see also Section \ref{subsec:sqzcoh_sqzvac}) and the coherent plus Fock (see also Section \ref{subsec:coh_plus_Fock_input}) input states.}. 

We thus assume $\braket{\hat{a}_0}=0$ and from equation \eqref{eq:VASP_shorthand_notations} we have $S_\pm=P=0$ implying
\begin{equation}
C'_2 = C'_4 = 0.
\end{equation}
A) If $C'_3=0$, too, then:
\begin{enumerate}
	\item[i)] if $C'_1<0$ then the optimum is in the degenerate case \emph{i. e.} $T^{(i)}_{opt}=0/1$ and $\mathcal{F}^{(i)}_{max}=C'_0$.
	\item[ii)] if $C'_1>0$ then the optimum is again in the balanced case, \emph{i. e.} $T^{(i)}_{opt}=1/\sqrt{2}$ and 
\begin{equation}
\mathcal{F}^{(i)}_{max}=C'_0+\frac{C'_1}{4}.
\end{equation}
\end{enumerate}
B) For the general case with $\{C'_1,\:C'_3\}\neq0$, we have the scenarios:
\begin{enumerate}
	\item[i)] if the constraints
\begin{equation}
\label{eq:T_i_opt_existence_cond_C2prime_C4prime_ZERO}
\left\{
\begin{array}{l}
A\geq8\left(\Delta^2\hat{n}_1-\text{Cov}(\hat{n}_0,\hat{n}_1)\right)\\
A\geq8\left(\Delta^2\hat{n}_0-\text{Cov}(\hat{n}_0,\hat{n}_1)\right)
\end{array}
\right.
\end{equation}
are met, then $0\leq T^{(i)}_{opt}\leq1$ exists and its value is
\begin{equation}
\label{eq:T_i_opt_C2prime_C4prime_ZERO}
T^{(i)}_{opt}
=\sqrt{\frac{1}{2}+\frac{C'_3}{C'_1}}
\end{equation}
yielding the maximum QFI,
\begin{equation}
\label{eq:F_i_in_shorthand_C_prime_C2prime_C4prime_ZERO_MAX}
\mathcal{F}^{(i)}_{max}
=C'_0+C'_1\left(\frac{1}{4}
+\frac{{C'_3}^2}{{C'_1}^2}
\right).
\end{equation}
	\item[ii)] if the conditions
\begin{equation}
\label{eq:T_i_opt_is_0_COND_C2prime_C4prime_ZERO}
\left\{
\begin{array}{l}
A\geq8\left(\Delta^2\hat{n}_1-\text{Cov}(\hat{n}_0,\hat{n}_1)\right)\\
A<8\left(\Delta^2\hat{n}_0-\text{Cov}(\hat{n}_0,\hat{n}_1)\right)
\end{array}
\right.
\end{equation}
are met then the optimal transmission coefficient is in the degenerate case ($T^{(i)}_{opt}=0$) and  
\begin{equation}\mathcal{F}^{(i)}_{max}=C'_0-C'_3=4\left(\Delta^2\hat{n}_0+\text{Cov}(\hat{n}_0,\hat{n}_1)\right).
\end{equation}
	\item[iii)] if the conditions
\begin{equation}
\label{eq:T_i_opt_is_1_COND_C2prime_C4prime_ZERO}
\left\{
\begin{array}{l}
A<8\left(\Delta^2\hat{n}_1-\text{Cov}(\hat{n}_0,\hat{n}_1)\right)\\
A\geq8\left(\Delta^2\hat{n}_0-\text{Cov}(\hat{n}_0,\hat{n}_1)\right)
\end{array}
\right.
\end{equation}
are met, then the optimal transmission coefficient is again in the degenerate case ($T^{(i)}_{opt}=1$) and 
\begin{equation}
\mathcal{F}^{(i)}_{max}
=C'_0+C'_3=4\left(\Delta^2\hat{n}_1+\text{Cov}(\hat{n}_0,\hat{n}_1)\right).
\end{equation}
	\item[iv)] finally, if
\begin{equation}
\label{eq:T_i_opt_is_0or1_COND_C2prime_C4prime_ZERO}
\left\{
\begin{array}{l}
A<8\left(\Delta^2\hat{n}_1-\text{Cov}(\hat{n}_0,\hat{n}_1)\right)\\
A<8\left(\Delta^2\hat{n}_0-\text{Cov}(\hat{n}_0,\hat{n}_1)\right)
\end{array}
\right.
\end{equation}
then then the optimal transmission coefficient is in the degenerate yielding $T^{(i)}_{opt}=0$ if $C'_3<0$ and $T^{(i)}_{opt}=1$ if $C'_3>0$.
\end{enumerate}
The second easily solvable scenario happens when $C_1'=C_2'=0$ (relevant for example for a double coherent input) and we get
\begin{equation}
\label{eq:T_i_opt_C1prime_C2prime_ZERO}
T^{(i)}_{opt}
=\sqrt{\frac{1}{2}+\frac{|C'_3|\text{sgn}(C'_3)}{\sqrt{4(C'_3)^2+(C'_4)^2}}}
\end{equation}
valid if $C'_4>0$ and the maximum QFI is given by
\begin{equation}
\label{eq:F_i_in_shorthand_C_prime_C1prime_C2prime_ZERO_MAX}
\mathcal{F}^{(i)}_{max}
=C'_0
+\frac{\sqrt{4(C_3')^2+(C_4')^2}}{2}.
\end{equation}
For $C'_4<0$ we have the optimum QFI in degenerate case with
\begin{equation}
\label{eq:T_i_opt_C1prime_C2prime_ZERO_degenerate}
T^{(i)}_{opt}
=\left\{
\begin{array}{lc}
0 & \textrm{if }C'_3<0\\
1 & \textrm{if }C'_3>0.\\
\end{array}
\right.
\end{equation}
The optimal transmission coefficient in the general case (when none of the $C'$ coefficients is assumed null) can be found in the form
\begin{equation}
\label{eq:T_opt_for_F_i_general_case}
T^{(i)}_{opt}=\sqrt{\frac{1\pm\sqrt{1-4\chi^2_{sol}}}{2}}
\end{equation}
where $\chi_{sol}$ are those solutions of the quartic equation \eqref{eq:app:F_i_quartic_equations} that obey $\chi_{sol}\in\mathbb{R}$ and $|\chi^2_{sol}|\leq0.5$. More details are found in Appendix \ref{sec:app:calc_T_i_opt_for_Fi}. The results from this section are summarized in Table~\ref{tab:T_i_and_F_i}.

\subsection{The symmetric single parameter QFI $\mathcal{F}^{(ii)}$}
\label{subsec:F_ii_optimal_T}
The symmetric single-parameter QFI \eqref{eq:Fii_variance_n3_minus_n2} can be written as
\begin{equation}
\label{eq:F_ii_shorthand}
\mathcal{F}^{(ii)}=C''_0
+C''_1\vert{TR}\vert^2
+C''_2|TR|(|T|^2-|R|^2)
\end{equation}
and the coefficients are given by
\begin{equation}
\label{eq:F_ii_C_sec_coeffs}
\left\{
\begin{array}{l}
C''_0=V_{+}-V_{cov}\\
C''_1=A-4(V_{+}-V_{cov})\\
C''_2=-2S_{+}.
\end{array}
\right.
\end{equation}
Since $\mathcal{F}^{(ii)}$ from equation \eqref{eq:F_ii_shorthand}  is formally identical to $\mathcal{F}^{(2p)}$ from equation \eqref{eq:F_2p_shorthand}, in the following we will employ the solutions from Section \ref{subsec:Fisher_2p} by simply replacing the $C$-coefficients with the corresponding $C''$-ones.

All results from this section are summarized in Table~\ref{tab:T_ii_and_F_ii}.

\begin{table}
\hspace*{-0.5cm}
\centering
\renewcommand{\arraystretch}{1.3}
\begin{tabular}{c||c|c|c|c|}
\arrayrulecolor{black}
\cline{2-5}
 & \multicolumn{4}{c|}{constraints obeyed by the $C''$-coefficients} \\
\cline{2-5}
\cline{2-5}
 & $C''_1=0$ & $C''_1\neq0$ & \multicolumn{2}{c|}{$C''_2=0$} \\
\cline{4-5}
 & $C''_2=0$ & $C''_2\neq0$ &  $C''_1>0$ & $C''_1<0$ \\
\arrayrulecolor{black}
\hline
\multicolumn{1}{|c||}{$T^{(ii)}_{opt}$} & irrelevant & $\sqrt{\frac{1+\text{sgn}(C''_2)\sqrt{\frac{1}{2}-\frac{\text{sgn}(C''_1)|C''_1|}{2\sqrt{{C''_1}^2+4{C''_2}^2}}}}{2}
}$ & $\frac{1}{\sqrt{2}}$   & $0/1$ \\
\hline
\multicolumn{1}{|c||}{$\mathcal{F}^{(ii)}_{max}$} & $C''_0$ & $C''_0+\frac{C''_1}{8}
+\frac{\sqrt{{C''_1}^2+4{C''_2}^2}}{8}$ &  $C''_0+\frac{C''_1}{4}$   & $C''_0$ \\
\hline
\end{tabular}
\caption{\label{tab:T_ii_and_F_ii}Optimal transmission coefficient $T^{(ii)}_{opt}$ and the corresponding maximum symmetric single-parameter QFI $\mathcal{F}^{(ii)}_{max}$ in all discussed scenarios.}
\end{table}


\section{Two noteworthy scenarios}
\label{sec:special_cases}
Before discussing the applications of the previous results to some interesting input states, we focus on two special situations. 

\subsection{The condition for no metrological advantage of an external phase reference}
\label{subsec:no_metrological_advantage_ext_phase}
From equation \eqref{eq:F_i_is_Fss_plus_Fdd-2Fds} (or equation \eqref{eq:F_i_n2_is_Fss_plus_Fdd_plus_2Fds} as a matter of fact) we can immediately obtain $\mathcal{F}^{(i)}\geq\mathcal{F}^{(2p)}$ and the condition $\mathcal{F}^{(ii)}\geq\mathcal{F}^{(2p)}$ is also straightforward from equations \eqref{eq:Fisher_information_F_2p_DEFINITION} and \eqref{eq:Fii_variance_n3_minus_n2}. In other words, from a quantum metrological point of view, \emph{having access to an external phase reference can only be beneficial}. Thus, an interesting question to answer would be the following: what input states render the availability of this external phase reference useless, irrespective of the transmission coefficient, $T$?

We start our discussion with the asymmetric single-parameter QFI.  The condition for having no metrological advantage with an external phase reference translates into ${\mathcal{F}^{(i)}=\mathcal{F}^{(2p)}}$, an equality that must be valid for any value of $T$. From equations \eqref{eq:F_2p_shorthand} and \eqref{eq:F_i_in_shorthand_C_prime} we immediately have 
\begin{equation}
\label{eq:No_metrological_advantage_Fi}
\left\{
\begin{array}{l}
V_++V_{cov}=0\\
V_-=0\\
S_-+P=0
\end{array}
\right.
\end{equation}
and if the input state is separable, the first two conditions morph into
\begin{equation}
\label{eq:No_metrological_advantage_Variance_n0_Variance_n1_equal_0}
\Delta^2\hat{n}_0=\Delta^2\hat{n}_1=0.
\end{equation}

In the case of the symmetric single-parameter QFI, the condition for no metrological advantage while having access to an external phase reference translates into $\mathcal{F}^{(ii)}=\mathcal{F}^{(2p)}$, implying immediately $\mathcal{F}_{sd}=0$. From equation \eqref{eq:F_sd_unbalanced_entangled_final} this condition imposes
\begin{equation}
\label{eq:No_metrological_advantage_Fii}
\left\{
\begin{array}{l}
V_-=0\\
S_-+P=0.
\end{array}
\right.
\end{equation}
This time the first condition (equally valid for entangled and separable input states) implies $\Delta^2{\hat{n}_0}-\Delta^2\hat{n}_1=0$, a much weaker constraint wrt equation \eqref{eq:No_metrological_advantage_Variance_n0_Variance_n1_equal_0}. 

We remark that if ${\mathcal{F}^{(i)}=\mathcal{F}^{2p}}$, then necessarily $\mathcal{F}^{(ii)}=\mathcal{F}^{2p}$, too. The converse is obviously not true.


As a first example we consider the twin-Fock input state,
\begin{equation}
\label{eq:psi_in_twin_Fock}
\ket{\psi_{in}}=\ket{n_1m_0}=\frac{(\hat{a}_1^\dagger)^n}{\sqrt{n!}}\frac{(\hat{a}_0^\dagger)^m}{\sqrt{m!}}\ket{0}
\end{equation}
and if we set $m=n$ it is also called a Holland-Burnett state \cite{Hol93}. Since the constraints \eqref{eq:No_metrological_advantage_Fi} are fulfilled, we have
\begin{equation}
\label{eq:F_2p_TR_equation_twin_Fock}
\mathcal{F}^{(2p)}=\mathcal{F}^{(i)}=\mathcal{F}^{(ii)}=4\vert{TR}\vert^2\left(n+m+2mn\right)
\end{equation}
and there is no metrological advantage in having an external phase reference for both single-parameter QFI scenarios.

\begin{figure}
	\includegraphics[scale=0.5]{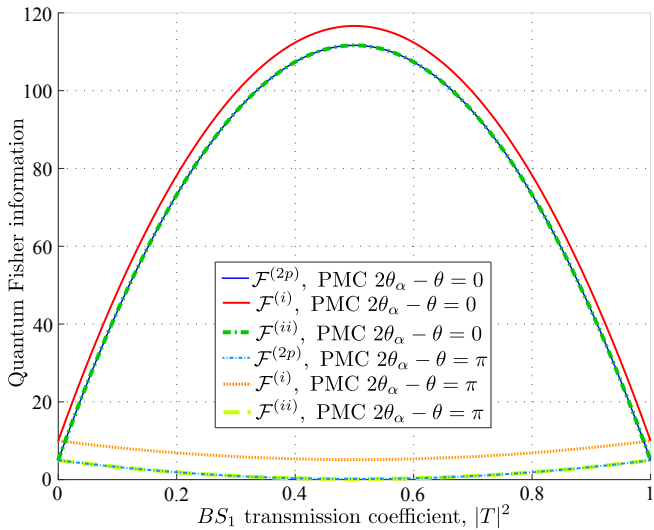}
	\caption{The three QFIs for a coherent plus squeezed vacuum input. Since our parameters obey the conditions \eqref{eq:No_metrological_advantage_Fii}, there is no metrological advantage in having an external phase reference for $\mathcal{F}^{(ii)}$. It is noteworthy that the equality $\mathcal{F}^{(2p)}=\mathcal{F}^{(ii)}$ remains valid for any value of the the input PMC, $2\theta_\alpha-\theta$. Parameters used: $r=1.9$ and $\vert\alpha\vert=\sinh2r/\sqrt{2}$.}
	\label{fig:Coh_sqz_vac_no_metrological_adv_Fii}
\end{figure}

As a second example we consider the rather popular coherent plus squeezed vacuum input,
\begin{equation}
\label{eq:psi_in_coh_sqzvac}
\vert\psi_{in}\rangle=\vert\alpha_1\xi_0\rangle
\end{equation}
where the coherent state in port $1$, $\ket{\alpha_1}=\hat{D}_1\left(\alpha\right)\ket{0_1}$, is obtained by applying the displacement or Glauber operator \cite{GerryKnight,MandelWolf,Aga12},
\begin{equation}
\label{eq:displacement_operator_def}
\hat{D}_1\left(\alpha\right)=e^{\alpha\hat{a}_1^\dagger-\alpha^*\hat{a}_1}
\end{equation}
with $\alpha=\vert\alpha\vert e^{i\theta_\alpha}$. The squeezed vacuum in port $0$ is obtained by applying the squeezing operator \cite{GerryKnight,Yue76}
\begin{equation}
\label{eq:Squeezing_operator}
\hat{S}_0\left(\xi\right)=e^{\frac{1}{2}\left[\xi^*\hat{a}_0^2-\xi(\hat{a}_0^\dagger)^2\right]}
\end{equation}
to the vacuum state, \emph{i. e.} ${\ket{\xi_0}=\hat{S}_0\left(\xi\right)\vert0_0\rangle}$. Here $\xi=re^{i\theta}$. Usually $r\in\mathbb{R}^{+}$ is called the squeezing factor and ${\theta}$ denotes the phase of the squeezed state.

Since the constraints \eqref{eq:No_metrological_advantage_Variance_n0_Variance_n1_equal_0} are impossible to satisfy with the input state from equation \eqref{eq:psi_in_coh_sqzvac}, we can try to satisfy the weaker constraints \eqref{eq:No_metrological_advantage_Fii}. Indeed if we choose now $\vert\alpha\vert$ and $r$ so that
\begin{equation}
\label{eq:no_metrological_adv_Fii_coh_sqz_vac}
\vert\alpha\vert=\frac{\sinh2r}{\sqrt{2}},
\end{equation}
the constraints from equation \eqref{eq:No_metrological_advantage_Fii} are satisfied. This scenario is depicted in Fig.~\ref{fig:Coh_sqz_vac_no_metrological_adv_Fii} for the parameter $r=1.9$. Although the QFI implied by the input state \eqref{eq:psi_in_coh_sqzvac} is heavily PMC-dependent \cite{Ata18,Ata20}, when condition \eqref{eq:no_metrological_adv_Fii_coh_sqz_vac} is fulfilled, there is no metrological advantage for $\mathcal{F}^{(ii)}$, regardless of the input PMC.

\subsection{Optimal phase sensitivity with one input in the vacuum state}
\label{subsec:optimality_one_input_vacuum}
Another interesting scenario arises with an interferometer having  one input in the vacuum state \cite{Takeoka2017}. Without loss of generality we choose the input port $0$ as ``dark'', \emph{i. e.} $\langle{\hat{n}_0}\rangle=\Delta^2{\hat{n}_0}=0$. From definitions \eqref{eq:VASP_shorthand_notations} we have
\begin{equation}
\label{eq:VASP_shorthand_notations_port0_vacuum}
\left\{
\begin{array}{l}
V_\pm=\pm\Delta^2{\hat{n}_1}\\
A=4\langle{\hat{n}_1}\rangle
\\
S_{\pm} = P = 0\\
\end{array}
\right.
\end{equation}
we thus get the Fisher matrix elements
\begin{equation}
\label{eq:F_ss_F_dd_F_sd_one_input_vacuum}
\left\{
\begin{array}{l}
\mathcal{F}_{ss}=\Delta^2\hat{n}_1\\
\mathcal{F}_{dd}=\Delta^2\hat{n}_1
+4\vert{TR}\vert^2(\langle{\hat{n}_1}\rangle-\Delta^2{\hat{n}_1})\\
\mathcal{F}_{sd}=-\left(\vert{T}\vert^2-\vert{R}\vert^2\right)
\Delta^2{\hat{n}_1}
\end{array}
\right.
\end{equation}
 The two parameter QFI \eqref{eq:Fisher_information_F_2p_DEFINITION} becomes
\begin{equation}
\label{eq:F_2p_one_input_vacuum}
\mathcal{F}^{(2p)}=4\vert{TR}\vert^2\langle{\hat{n}_1}\rangle
\end{equation}
and two conclusions are immediate:
\begin{enumerate}
	\item[i)] the QFI $\mathcal{F}^{(2p)}$ is maximal in the balanced case
	\item[ii)] the phase sensitivity cannot surpass the SQL \cite{Takeoka2017}
\end{enumerate}
In the case of the asymmetric single-parameter QFI we get $C_0'=C_3'=2\Delta^2{\hat{n}_1}$, $C_1'=4(\langle{\hat{n}_1}\rangle-\Delta^2{\hat{n}_1})$ and $C'_2=C'_4=0$ implying
\begin{eqnarray}
\mathcal{F}^{(i)}
=4\langle{\hat{n}_1}\rangle\vert{T}\vert^2
-4(\langle{\hat{n}_1}\rangle-\Delta^2{\hat{n}_1})\vert{T}\vert^4
\end{eqnarray}
Once again we consider $T$ real and conclude that:
\begin{enumerate}
	\item[i)] for $\Delta^2{\hat{n}_1}\geq\frac{\langle{\hat{n}_1}\rangle}{2}$ the optimal transmission coefficient is $T^{(i)}_{opt}=1$, yielding the maximal QFI $\mathcal{F}^{(i)}=4\Delta^2{\hat{n}_1}$.
	\item[ii)] for $\Delta^2{\hat{n}_1}<\frac{\langle{\hat{n}_1}\rangle}{2}$ the optimal transmission coefficient is
	\begin{equation}
\label{eq:T_i_opt_C2prime_C4prime_ZERO_one_input_vacuum}
T^{(i)}_{opt}=\sqrt{\frac{\langle{\hat{n}_1}\rangle}
{2\left(\langle{\hat{n}_1}\rangle-\Delta^2{\hat{n}_1}\right)}}.
\end{equation}
and it implies the maximum QFI,
\begin{equation}
\label{eq:Fi_max_port0_Vacuum}
\mathcal{F}^{(i)}_{max}=\frac{\braket{\hat{n}_1}^2}{\langle{\hat{n}_1}\rangle-\Delta^2{\hat{n}_1}}.
\end{equation}
\end{enumerate}
We remark that the above-mentioned limit ($\Delta^2{\hat{n}_1}<{\langle{\hat{n}_1}\rangle}/{2}$) is simply the existence condition  \eqref{eq:T_i_opt_existence_cond_C2prime_C4prime_ZERO} for $T^{(i)}_{opt}$ adapted when port $0$ is in the vacuum state.

For the symmetric single-parameter QFI $\mathcal{F}^{(ii)}=\mathcal{F}_{dd}$ and from equation \eqref{eq:F_ss_F_dd_F_sd_one_input_vacuum} we have the coefficients $C''_0=\Delta^2{\hat{n}_1}$, $C''_1=4(\langle{\hat{n}_1}\rangle-\Delta^2{\hat{n}_1})$ and $C''_2=0$. The optimum transmission coefficient maximizing this QFI is
\begin{equation}
\label{eq:T_opt_squared_F_ii_one_input_vacuum}
{T}_{opt}^{(ii)}
=
\left\{
\begin{array}{ll}
0/1 & \text{ if } \langle{\hat{n}_1}\rangle<\Delta^2{\hat{n}_1}\\
irrelevant & \text{ if } \langle{\hat{n}_1}\rangle=\Delta^2{\hat{n}_1}\\
\frac{1}{\sqrt{2}} & \text{ if } \langle{\hat{n}_1}\rangle>\Delta^2{\hat{n}_1}
\end{array}
\right.
\end{equation}
and we considered $T\in\mathbb{R}$ again for simplicity. Thus, if the input state at port $1$ has a Poissonian statistics ${T}_{opt}$ is irrelevant since the QFI is constant, $\mathcal{F}^{(ii)}=\Delta^2{\hat{n}_1}=\braket{\hat{n}_1}$. This results carries on even if both inputs have Poissonian statistics \cite{Ata20}.

We provide in the following three examples of growing complexity allowing us to apply all aforementioned results.
%
%
As a first example we employ the single Fock input state \emph{i. e.} equation \eqref{eq:psi_in_twin_Fock} with $m=0$,
\begin{equation}
\label{eq:psi_in_single_Fock_input}
\ket{\psi_{in}}=\ket{n_10_0}.
\end{equation}
From equation \eqref{eq:F_2p_TR_equation_twin_Fock}, the two-parameter QFI (see the discussion in Appendix \ref{sec:app:single_Fock_calculations}) is
\begin{equation}
\label{eq:F_2p_single_Fock_input}
\mathcal{F}^{(2p)}=4|TR|n.
\end{equation}
Since the conditions from equation \eqref{eq:No_metrological_advantage_Fi} are satisfied we have
\begin{equation}
\mathcal{F}^{(2p)}=\mathcal{F}^{(i)}=\mathcal{F}^{(ii)}=4|TR|n.
\end{equation}
thus $T_{opt}^{(2p)}=T_{opt}^{(i)}=T_{opt}^{(ii)}=1/\sqrt{2}$ and $\mathcal{F}_{max}^{(2p)}=\mathcal{F}_{max}^{(i)}=\mathcal{F}_{max}^{(ii)}=n$.  The fact that there is no metrological advantage in having an external phase reference for a single Fock input can be seen as a quantum metrological proof that Fock states do not have a well defined phase \cite{GerryKnight}.

\begin{figure}
		\includegraphics[scale=0.5]{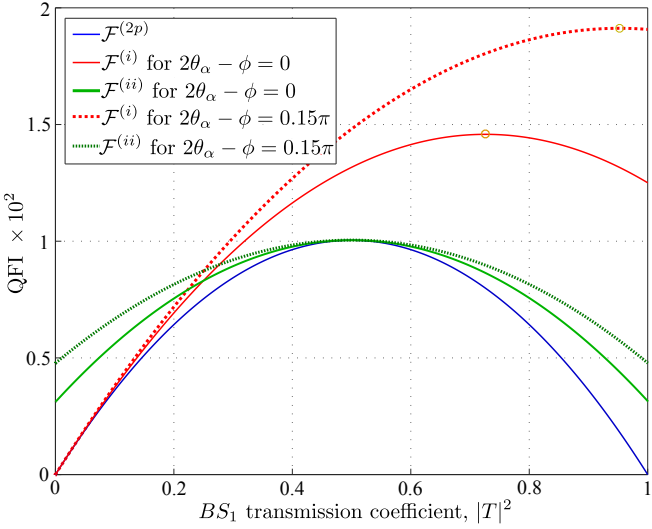}
	\caption{The three QFIs for a squeezed-coherent state applied to input port $1$ with the second input port in vacuum for two phase-matching conditions. Parameters used: $\vert\alpha\vert=10$, $z=0.6$. The circles mark the maxima for the two $\mathcal{F}^{(i)}$ curves.}
	\label{fig:Fisher_CohSqz1_Vac0_phi_0_015pi}
\end{figure}


As a second example we consider the single coherent input, $\vert\psi_{in}\rangle=\vert\alpha_10_0\rangle$. Since this scenario was already discussed in reference \cite{Ata20}, we simply connect the results to the formalism of this paper. In the two-parameter scenario we have $T_{opt}^{(2p)}=1/\sqrt{2}$ and $\mathcal{F}_{max}^{(2p)}=\vert\alpha\vert^2$. Since $\Delta^2\hat{n}_1>\braket{\hat{n}_1}/2$, for the asymmetric single parameter QFI we have optimality for $T_{opt}^{(i)}=1$ and $\mathcal{F}_{max}^{(i)}=4\vert\alpha\vert^2$. For the symmetric single parameter QFI, from equation \eqref{eq:T_opt_squared_F_ii_one_input_vacuum} we get that $T_{opt}^{(ii)}$ is irrelevant and $\mathcal{F}_{max}^{(ii)}=\vert\alpha\vert^2$. For the asymmetric single-parameter QFI we always have $\mathcal{F}^{(i)}\geq\mathcal{F}_{max}^{(2p)}$. Finally, for the symmetric single parameter QFI we have $\mathcal{F}^{(ii)}>\mathcal{F}^{(2p)}$ for any $T\neq1/\sqrt{2}$. Thus, having access to an external phase reference brings a clear advantage for both $\mathcal{F}^{(i)}$ and $\mathcal{F}^{(ii)}$.

\begin{figure}
	\includegraphics[scale=0.5]{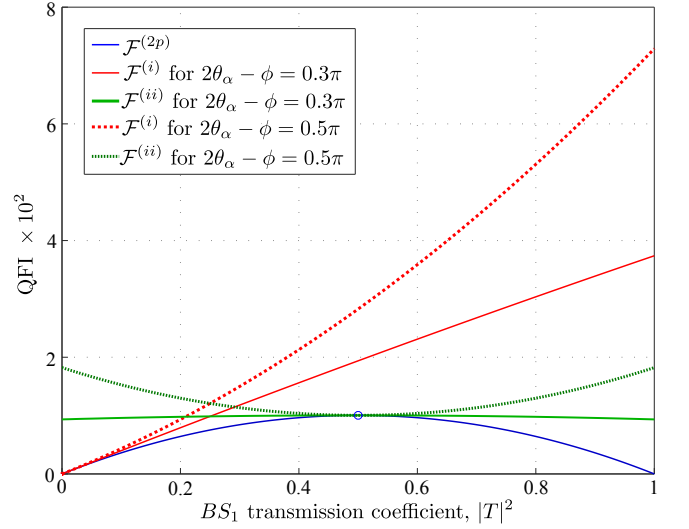}
	\caption{The three QFIs for a squeezed-coherent state applied to input port $1$ with the second input port in vacuum for two phase-matching conditions. Parameters used: $\vert\alpha\vert=10$, $z=0.6$. The circle marks the maximum for the $\mathcal{F}^{(2p)}$ curve.}
	\label{fig:Fisher_CohSqz1_Vac0_phi_03_05pi}
\end{figure}


A final and more complex input state that can depict all scenarios described at the beginning of this section is the squeezed-coherent plus vacuum input state,
\begin{equation}
\label{eq:psi_in_sqz-coh_vaccum0}
\vert\psi_{in}\rangle=\vert(\alpha\zeta)_10_0\rangle
\end{equation}
where $\ket{(\alpha\zeta)_1}=\hat{D}_1\left(\alpha\right)\hat{S}_1\left(\zeta\right)\vert0_1\rangle$ and $\zeta=ze^{i\phi}$. We have an average number of input photons 
\begin{equation}
\label{eq:Average_n1_sqz_coh_VACUUM0}
\langle{\hat{n}_1}\rangle=\vert\alpha\vert^2+\sinh^2z
\end{equation}
and a variance
\begin{eqnarray}
\label{eq:Variance_n1_sqz_coh_plus_sqz_vac_VACUUM0}
\Delta^2\hat{n}_1=\frac{\sinh^22z}{2}
+{\vert\alpha\vert^2}\left(\cosh2z
\right.
\nonumber\\
\left.
-\sinh2z\cos\left(2\theta_\alpha-\phi\right)\right)
\end{eqnarray}
adjustable via the input PMC, $2\theta_\alpha-\phi$. Indeed, setting $2\theta_\alpha-\phi=0$ implies $\Delta^2{\hat{n}_1}={\sinh^22z}/{2}+{\vert\alpha\vert^2}e^{-2z}$ and this is the minimum variance one can achieve with this input state. On the contrary, setting $2\theta_\alpha-\phi=\pm\pi$ yields $\Delta^2{\hat{n}_1}={\sinh^22z}/{2}+{\vert\alpha\vert^2}e^{2z}$ and this time the input state \eqref{eq:psi_in_sqz-coh_vaccum0} yields its maximal variance.

In Fig.~\ref{fig:Fisher_CohSqz1_Vac0_phi_0_015pi} we depict two scenarios when $\Delta^2{\hat{n}_1}\leq\braket{\hat{n}_1}/2$. The two-parameter QFI $\mathcal{F}^{(2p)}$ (blue solid curve) having no dependence on $\Delta^2{\hat{n}_1}$ implies that the two-parameter QFI is invariant wrt the input PMC. Thus, for both scenarios, $\mathcal{F}^{(2p)}$ reaches its maximum $\mathcal{F}^{(2p)}_{max}=\langle{\hat{n}_1}\rangle$ for the balanced case, this statement remaining true for the two scenarios depicted in Fig.~\ref{fig:Fisher_CohSqz1_Vac0_phi_03_05pi}. 

The other two solid lines from {Fig.~\ref{fig:Fisher_CohSqz1_Vac0_phi_0_015pi} depict $\mathcal{F}^{(i)}$ and $\mathcal{F}^{(ii)}$
for the input PMC $2\theta_\alpha-\phi=0$. Since $\Delta^2{\hat{n}_1}<\langle{\hat{n}_1}\rangle$,  the single-parameter symmetric QFI $\mathcal{F}^{(ii)}$ is maximized in the balanced case while $\mathcal{F}^{(i)}$ peaks at
 ${T}_{opt}^{(i)}$ given by equation \eqref{eq:T_i_opt_C2prime_C4prime_ZERO_one_input_vacuum}. Given the parameters used (see the caption of Fig.~\ref{fig:Fisher_CohSqz1_Vac0_phi_0_015pi}) the optimum transmission coefficient \eqref{eq:T_i_opt_C2prime_C4prime_ZERO_one_input_vacuum} is found to be $T^{(i)}_{opt}\approx\sqrt{0.72}$. The second scenario depicted in Fig.~\ref{fig:Fisher_CohSqz1_Vac0_phi_0_015pi} (red dotted line) still  obeys $\Delta^2{\hat{n}_1}\leq\braket{\hat{n}_1}/2$, however this time the inequality is barely satisfied. Indeed, with the PMC $2\theta_\alpha-\phi=0.15\pi$ we find $T^{(i)}_{opt}=\sqrt{0.95}\approx1$ and thus  $\mathcal{F}^{(i)}_{max}\approx4\Delta^2{\hat{n}_1}$. Since $\Delta^2{\hat{n}_1}<\langle{\hat{n}_1}\rangle$ the single-parameter symmetric QFI is maximized in the balanced case.

In Fig.~\ref{fig:Fisher_CohSqz1_Vac0_phi_03_05pi} we depict two situations when condition $\Delta^2{\hat{n}_1}\leq\braket{\hat{n}_1}/2$ is no longer satisfied. Thus, the optimum transmission coefficient for the asymmetric single parameter QFI is ${T}_{opt}^{(i)}=1$ and $\mathcal{F}^{(i)}_{max}=4\Delta^2\hat{n}_1$ for both PMCs (red solid and, respectively, dotted curve). For input the PMC $\theta_\alpha-\phi=0.3\pi$ $\mathcal{F}^{(ii)}$ (green solid curve) although still peaked for $T^{(ii)}_{opt}=1/\sqrt{2}$, it is almost flat, making the very notion of optimal transmission coefficient less relevant.

For the input PMC $\theta_\alpha-\phi=0.5\pi$ we have $\Delta^2{\hat{n}_1}>\braket{\hat{n}_1}$ and from equation \eqref{eq:T_opt_squared_F_ii_one_input_vacuum} we find $T^{(ii)}_{opt}=0/1$, the symmetric single-parameter QFI, $\mathcal{F}^{(ii)}$, (green dotted curve) being thus maximized in the degenerate case.

\section{Gaussian and non-Gaussian input state examples}
\label{sec:examples}

A plethora of quantum states have been shown to have a quantum metrological interest \cite{Ata19,Cav81,Ralph2002,Anisimov2010,Joo2011,Birrittella2012,Carranza2012,
Birrittella2014,Ouyang2016,Hou2019,Wang2019,Par95,Spa15}. The discussions however, were carried out in the balanced case only, with few exceptions \cite{Ata20,Pre19,Zhong2020}. In this section we re-discuss a number of these states in the non-balanced scenario for all three QFIs.

\subsection{Squeezed-coherent plus squeezed vacuum input}
\label{subsec:sqzcoh_sqzvac}
Consider the squeezed-coherent plus squeezed vacuum input state \cite{Par95,Pre19,Ata20},
\begin{equation}
\label{eq:psi_in_sqzcoh_plus_sqz_vac}
\vert\psi_{in}\rangle=\vert(\alpha\zeta)_1\xi_0\rangle
\end{equation}
and we recall the notations for the two squeezers: ${\xi=re^{i\theta}}$ and ${\zeta=ze^{i\phi}}$. All QFIs are maximized if we impose the input PMC \cite{Ata20},
\begin{equation}
\label{eq:PMC_sqz-coh_plus_sqz-vac}
\left\{
\begin{array}{l}
2\theta_\alpha-\theta=0\\
2\theta_\alpha-\phi=\pm\pi.
\end{array}
\right.
\end{equation}
Calculations are detailed in Appendix \ref{sec:app:sqzcoh_sqzvac_calculations}. Since $C_1>0$ (due to the PMC choice) and $C_2=0$, equation \eqref{eq:T_opt_squared_F_2p} immediately implies that for the two-parameter QFI the optimum is found in the balanced case. 
%
\begin{figure}
	\includegraphics[scale=0.5]{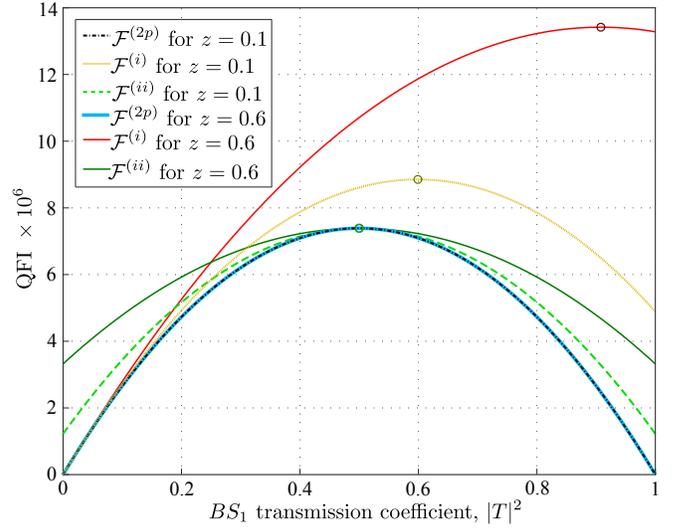}
	\caption{The three QFIs for a squeezed-coherent plus squeezed vacuum input state. Parameters used: $\vert\alpha\vert=10^3$, $r=1$, PMC $2\theta_\alpha-\theta=0$, $2\theta_\alpha-\phi=\pi$. The squeezing in port 1 is $z=0.1$ (dashed lines) and $z=0.6$ (solid lines). The circles mark the maximum value for each curve.}
	\label{fig:Sqzcoh_sqzvac_alpha10_r1_z_01_06_PMC_opt}
\end{figure}
For the asymmetric single parameter QFI, the optimum transmission coefficient $T^{(i)}_{opt}$ is found via equations \eqref{eq:F_i_C_prime_coeffs} and \eqref{eq:app:VASP_shorthand_notations_sqzcoh_sqzvac}. If one imposes the optimal input PMC \eqref{eq:PMC_sqz-coh_plus_sqz-vac}, $T^{(i)}_{opt}$  simplifies to the expression given in equation \eqref{eq:T_i_opt_C2prime_C4prime_ZERO_sqzcoh_sqz_vac_PMC}. In the  experimentally interesting high-intensity coherent regime \emph{i. e.} $\vert\alpha\vert^2\gg\{\sinh^2r,\:\sinh^2z\}$, from equation \eqref{eq:T_i_opt_C2prime_C4prime_ZERO_sqzcoh_sqz_vac_PMC} we can approximate \cite{Ata20},
\begin{equation}
\label{eq:T_opt_i_sqzcoh_sqzvac_high_alpha}
T^{(i)}_{opt}\approx
\sqrt{\frac{1}{2(1-e^{2(z-r)})}}
\end{equation}
and one can select an optimum transmission coefficient by adjusting the ratio of the squeezing factors. Since $C''_1>0$ and $C''_2=0$, the symmetric single parameter QFI is optimized for in the balanced case, too, thus ${T^{(ii)}_{opt}=1/\sqrt{2}}$.

In Fig.~\ref{fig:Sqzcoh_sqzvac_alpha10_r1_z_01_06_PMC_opt} we depict the three QFIs for a squeezed-coherent plus squeezed vacuum input state in the high-coherent regime (see figure caption for the parameters used). The dotted/dashed lines depict the case $z=0.1$, and the solid ones $z=0.6$. While the QFIs $\mathcal{F}^{(2p)}$ and $\mathcal{F}^{(ii)}$ are maximized in the balanced case (irrespective on the value of $z$), this is not true for $\mathcal{F}^{(i)}$. Increasing the squeezing factor $z$ has a notable effect on $\mathcal{F}^{(i)}$, and, remarkably, this advantage does not vanish in the experimentally interesting high-coherent regime, $\vert\alpha\vert^2\gg\{\sinh^2r,\:\sinh^2z\}$.

We conclude that the input state \eqref{eq:psi_in_sqzcoh_plus_sqz_vac} shows a metrological advantage if an external phase reference is available. In reference \cite{Ata20} it has been shown that the theoretically predicted QFI $\mathcal{F}^{(i)}$ can be approached via a balanced homodyne detection technique, by suitably adjusting the transmission coefficient of the second beam splitter, $BS_2$.

\subsection{Squeezed-coherent plus squeezed-coherent input}
\label{subsec:sqzcoh_sqzcol}
We consider now the squeezed-coherent plus squeezed-coherent input state \cite{Ata19},
\begin{equation}
\label{eq:psi_in_sqzcoh_plus_sqzcoh}
\vert\psi_{in}\rangle=\vert(\alpha\zeta)_1(\beta\xi)_0\rangle
\end{equation}
where for port $0$ we have $\ket{(\beta\xi)_0}=\hat{D}_0\left(\beta\right)\hat{S}_0\left(\xi\right)\vert0\rangle$ and $\beta=\vert\beta\vert e^{i\theta_\beta}$. Calculations are detailed in Appendix \ref{sec:app:sqzcoh_sqzcoh_calculations}. We remark that since this state does not necessarily imply $C_2=0$, the optimum for the two-parameter QFI is not always in the balanced case. Unless a reduced scenario is possible, the optimum transmission coefficient ${T}_{opt}^{(i)}$ is obtained from equation \eqref{eq:T_opt_for_F_i_general_case}.

\begin{figure}
	\includegraphics[scale=0.5]{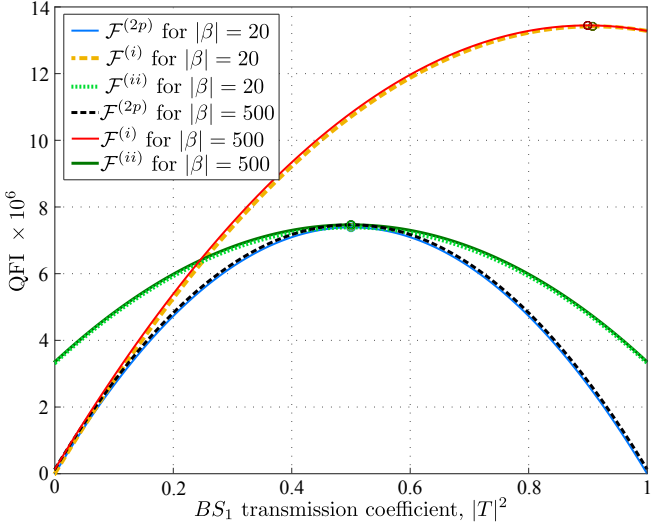}
	\caption{The three QFIs for a squeezed-coherent plus squeezed-coherent input state. The addition of the second coherent source brings no noticeable advantage. Parameters used: $\vert\alpha\vert=10^3$, $r=1$, $z=0.6$, PMC $2\theta_\alpha-\theta=0$, $2\theta_\alpha-\phi=\pi$ and $\theta_\alpha-\theta_\beta=0$. The circles mark the maximum for each curve.}
	\label{fig:SqzCoh1_SqzCoh0_alpha10_r1_z0_6_PMC1_beta_02_25_versus_T}
\end{figure}

In reference \cite{Ata19}, the two-parameter QFI for squeezed-coherent plus squeezed-coherent input state was thoroughly discussed in the balanced case. Among the input PMCs that maximize $\mathcal{F}^{(2p)}$, the first, denoted by  (PMC1)  involved the constraints from equation \eqref{eq:PMC_sqz-coh_plus_sqz-vac} plus the supplementary condition
\begin{equation}
\label{eq:theta_alpha_minus_theta_beta_ZERO}
\theta_\alpha-\theta_\beta=0
\end{equation}
for the second coherent source. From equation \eqref{eq:app:shorthand_VASP_sqzcoh_sqzcoh_PMC1} we can immediately deduce $C_2=0$ and $C''_2=0$, the QFIs $\mathcal{F}^{(2p)}$ and $\mathcal{F}^{(ii)}$ being thus maximized in the balanced case. Since $S_{\pm}=P=0$, we can use the first reduced scenario from Section \eqref{subsec:F_i_optimal_T} to find $T_{opt}^{(i)}$.
In Fig.~\ref{fig:SqzCoh1_SqzCoh0_alpha10_r1_z0_6_PMC1_beta_02_25_versus_T} we plot this scenario. We basically keep the same parameters used in Fig.~\ref{fig:Sqzcoh_sqzvac_alpha10_r1_z_01_06_PMC_opt} (the $z=0.6$ case is selected) and start increasing the coherent amplitude $\vert\beta\vert$ in port 1, while obeying to the PMC \eqref{eq:theta_alpha_minus_theta_beta_ZERO}.  As seen from  Fig.~\ref{fig:SqzCoh1_SqzCoh0_alpha10_r1_z0_6_PMC1_beta_02_25_versus_T}, the incrementation of the second coherent source from $\vert\beta\vert=20$ to $\vert\beta\vert=500$ brings an irrelevant increase for all QFIs. Further increasing $\vert\beta\vert\leq\vert\alpha\vert$ still yields a minimal increase for all considered QFIs. We conclude that with the PMC \eqref{eq:theta_alpha_minus_theta_beta_ZERO} all energy put into the coherent beam from port $0$ is simply wasted.


The second scenario, denoted (PMC2) \cite{Ata19},
\begin{equation}
\label{eq:PMC2}
\left\{
\begin{array}{l}
2\theta_\alpha-\theta=0\\
2\theta_\alpha-\phi=0\\
\theta_\alpha-\theta_\beta=0.
\end{array}
\right.
\end{equation}
was shown to be adapted to the high-coherent regime ($\{\vert\alpha\vert^2,\vert\beta\vert^2\}\gg\{\sinh^2r,\sinh^2z\}$), at least as far as $\mathcal{F}^{(2p)}$ is concerned. We extend now this scenario for all three QFIs. Since again from equations \eqref{eq:app:shorthand_VASP_sqzcoh_sqzcoh_PMC2} we find $C_2=C''_2=0$, we have $T_{opt}^{(2p)}=T_{opt}^{(ii)}=1/\sqrt{2}$. Finding again $S_{\pm}=P=0$, we can use the first reduced scenario from Section \eqref{subsec:F_i_optimal_T} to find $T_{opt}^{(i)}$.  We depict this scenario in Fig.~\ref{fig:Fig9_SqzCoh1_SqzCoh0_alpha1000_r1_z0_6_opt_beta_20_250_PMC2}. The same input parameters from Fig.~\ref{fig:SqzCoh1_SqzCoh0_alpha10_r1_z0_6_PMC1_beta_02_25_versus_T} are employed, except for the input PMCs. While $\mathcal{F}^{(2p)}$ and $\mathcal{F}^{(ii)}$ show a relative enhancement as $\vert\beta\vert$ increases, the performance of $\mathcal{F}^{(i)}$ remains modest, well below the results from Fig.~\ref{fig:SqzCoh1_SqzCoh0_alpha10_r1_z0_6_PMC1_beta_02_25_versus_T}. We conclude that (PMC2) is, too, a rather poor choice as far as the maximization of $\mathcal{F}^{(i)}$ is concerned.

\begin{figure}
	\includegraphics[scale=0.5]{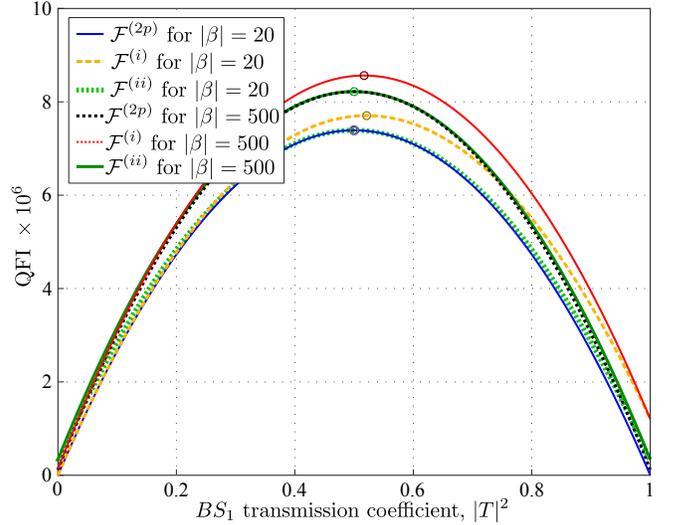}
	\caption{The three QFIs for a squeezed-coherent plus squeezed-coherent input state.  The addition of the second coherent source brings some increase for $\mathcal{F}^{(2p)}$ and $\mathcal{F}^{(ii)}$ in the case of $\mathcal{F}^{(i)}$.  Parameters used: $\vert\alpha\vert=10^3$, $r=1$, $z=0.6$, PMC $2\theta_\alpha-\theta=0$, $2\theta_\alpha-\phi=0$ and $\theta_\alpha-\theta_\beta=0$. Each circle marks the maximum of the corresponding curve.}
	\label{fig:Fig9_SqzCoh1_SqzCoh0_alpha1000_r1_z0_6_opt_beta_20_250_PMC2}
\end{figure}


We discuss now the last scenario, implying the PMCs from equation \eqref{eq:PMC_sqz-coh_plus_sqz-vac} plus the condition
\begin{equation}
\label{eq:theta_alpha_minus_theta_beta_PI_over_2}
\theta_\alpha-\theta_\beta=\frac{\pi}{2}
\end{equation}
and we denote these combined constraints by (PMC3). In reference \cite{Ata19}, employing the two-parameter QFI, $\mathcal{F}^{(2p)}$, (PMC3) has been shown to be optimal in the low coherent regime ($\{\vert\alpha\vert^2,\vert\beta\vert^2\}\ll\{\sinh^2r,\sinh^2z\}$). We will use it nonetheless in the high-coherent regime and show that, surprisingly, it can actually bring a substantial metrological advantage, however not for $\mathcal{F}^{(2p)}$, but for the single-parameter QFI, $\mathcal{F}^{(i)}$. From equations \eqref{eq:app:shorthand_VASP_sqzcoh_sqzcoh_PMC3} it becomes obvious that for this scenario, none of the involved QFIs is necessarily optimized in the balanced case.

The results for (PMC3) are depicted in Fig.~\ref{fig:Sqzcoh_sqzcoh_alpha10_r1_z06_PMC1_beta_01_1_2_5}. Even for a small coherent amplitude in port $0$ \emph{i. e.} $\vert\beta\vert=20$ ($\vert\beta\vert^2\ll\vert\alpha\vert^2$) its effect is noticeable when it comes to the single-parameter QFI, $\mathcal{F}^{(i)}$. Contrary to (PMC1) and (PMC2), $\mathcal{F}^{(i)}_{max}$ rapidly increases with $\vert\beta\vert$ and the (PMC3) scenario simply outperformes the previously discussed ones.


We conclude that employing the PMCs \eqref{eq:PMC_sqz-coh_plus_sqz-vac} and \eqref{eq:theta_alpha_minus_theta_beta_PI_over_2} for a squeezed-coherent plus squeezed-coherent input state is more than justified if an external phase reference is available. In this case, it can bring a real gain even for small values of the coherent amplitude $\vert\beta\vert$. Even more interestingly, this advantage remains intact in the experimentally interesting scenario where $\vert\alpha\vert^2\gg\{\sinh^2r,\:\sinh^2z\}$ and $\vert\beta\vert^2\gg\{\sinh^2r,\:\sinh^2z\}$.

In reference \cite{Spa15} the input state \eqref{eq:psi_in_sqzcoh_plus_sqzcoh} with a single-parameter QFI $\mathcal{F}^{(i)}$ was employed and the authors concluded that: ``Unbalanced devices may be also considered, which however lead to inferior performances''. As we showed in this section, some input parameters and PMCs confirm their assessment, however, others contradict it.

\begin{figure}
	\includegraphics[scale=0.5]{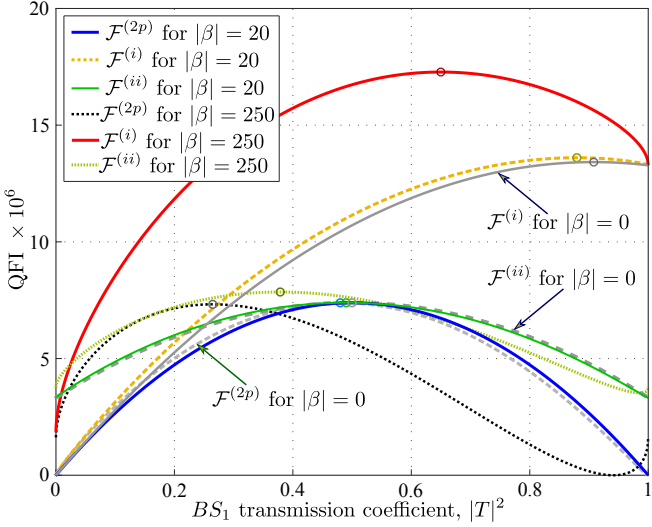}
	\caption{The three QFIs for a squeezed-coherent plus squeezed-coherent input state.  The addition of the second coherent source brings a hefty increase in the case of $\mathcal{F}^{(i)}$.  Parameters used: $\vert\alpha\vert=10^3$, $r=1$, $z=0.6$, PMC $2\theta_\alpha-\theta=0$, $2\theta_\alpha-\phi=\pi$ and $\theta_\alpha-\theta_\beta=\pi/2$. Each circle marks the maximum of the corresponding curve. The gray curves are given as reference and correspond to the ones from Fig.~\ref{fig:Sqzcoh_sqzvac_alpha10_r1_z_01_06_PMC_opt}, \emph{i. e.} for $\beta=0$.}
	\label{fig:Sqzcoh_sqzcoh_alpha10_r1_z06_PMC1_beta_01_1_2_5}
\end{figure}

\subsection{Coherent plus Fock input}
\label{subsec:coh_plus_Fock_input}
Consider the coherent plus Fock input state,
\begin{equation}
\label{eq:psi_in_Fock_plus_coherent}
\vert\psi_{in}\rangle=\vert\alpha_1n_0\rangle
\end{equation}
and from equation \eqref{eq:VASP_shorthand_notations} we have
\begin{equation}
\label{eq:VASP_shorthand_notations_Fock_plus_coherent}
\left\{
\begin{array}{l}
V_\pm=\pm\vert\alpha\vert^2\\
A=4\left(n+\vert\alpha\vert^2+2n\vert\alpha\vert^2\right)\\
S_\pm = P =0
\end{array}
\right.
\end{equation}
The $C$-coefficients ${C_0=C_2=0}$ and $C_1=4\left(n+2n\vert\alpha\vert^2\right)$ are immediately obtained from equation \eqref{eq:F_2p_C_coefficients_entangled}, the two-parameter QFI is thus found to be
\begin{equation}
\label{eq:F_2p_TR_Coh_plus_Fock}
\mathcal{F}^{(2p)}=\vert\alpha\vert^2+4|TR|^2\left(n+2n\vert\alpha\vert^2\right).
\end{equation}
Since $C_1>0$ ($C_1=0$ only if $\alpha=0$ and $n=0$, \emph{i. e.} the input is the vacuum state) the optimum transmission coefficient is $T_{opt}^{(2p)}=1/\sqrt{2}$ and in this case we get the maximum two-parameter QFI 
\begin{equation}
\label{eq:F_2p_TR_Coh_plus_Fock_maximum}
\mathcal{F}^{(2p)}_{max}=\vert\alpha\vert^2+n(1+2\vert\alpha\vert^2),
\end{equation}
result also reported in reference \cite{Birrittella2021}. The $C'$-coefficients are found to be
\begin{equation}
\label{eq:F_i_C_prime_coeffs_Fock_plus_coherent}
\left\{
\begin{array}{l}
C_0'=C_3'=2\vert\alpha\vert^2\\
C_1'=4n\left(1+2\vert\alpha\vert^2\right)\\
C_2'=C_4'=0
\end{array}
\right.
\end{equation}
consequently the asymmetric single-parameter QFI reads
\begin{equation}
\label{eq:F_i_Fock_plus_coherent}
\mathcal{F}^{(i)}=2\vert\alpha\vert^2
+4n\left(1+2\vert\alpha\vert^2\right)\vert{TR}\vert^2
+2\vert\alpha\vert^2(|T|^2-|R|^2).
\end{equation}
Imposing a balanced interferometer yields 
\begin{equation}
\label{eq:F_i_Fock_plus_coherent_BALANCED}
\mathcal{F}^{(i)}=2\vert\alpha\vert^2
+n\left(1+2\vert\alpha\vert^2\right)
\end{equation}
however, this value is not optimal\footnote{In reference \cite{Birrittella2012} the value reported for the QFI is $\mathcal{F}^{(i)}=4(\vert\alpha\vert^2
+n\left(1+2\vert\alpha\vert^2\right)\cos\varphi)$ and curiously, their result depends on the internal phase shift, $\varphi$, although being the result of the unitary generator $\hat{G}=\hat{J}_z$, it shouldn't \cite{Paris2009}. If we consider the best case scenario, \emph{i. e.} $\varphi\to0$, we still get four times $
\mathcal{F}^{(2p)}$ from equation \eqref{eq:F_2p_TR_Coh_plus_Fock_maximum}, not $\mathcal{F}^{(i)}$ from equation \eqref{eq:F_i_Fock_plus_coherent_BALANCED}.}. The optimum transmission coefficient $T^{(i)}_{opt}$ is given by equation \eqref{eq:T_i_opt_C2prime_C4prime_ZERO} and for this scenario is found to be
\begin{equation}
\label{eq:T_i_opt_C2prime_C4prime_ZERO_Fock_plus_coherent}
T^{(i)}_{opt}
=\sqrt{\frac{1}{2}+\frac{\vert\alpha\vert^2}{2n\left(1+2\vert\alpha\vert^2\right)}}
\end{equation}
result that is valid\footnote{For $n=0$ the input state \eqref{eq:psi_in_Fock_plus_coherent} degenerates into a single coherent input state \eqref{eq:psi_in_single_Fock_input} and the discussion from Section \ref{subsec:optimality_one_input_vacuum} applies.} for $n\neq0$. Inserting this value into equation \eqref{eq:F_i_Fock_plus_coherent} yields the maximum QFI,
\begin{equation}
\label{eq:F_i_Fock_plus_coherent_MAX}
\mathcal{F}^{(i)}_{max}
=2\vert\alpha\vert^2+n\left(1+2\vert\alpha\vert^2\right)+\frac{\vert\alpha\vert^4}
{n\left(1+2\vert\alpha\vert^2\right)}
\end{equation}
For the high coherent input regime ($\vert\alpha\vert^2\gg n$, $\vert\alpha\vert^2\gg 1$) we can approximate 
\begin{equation}
\mathcal{F}^{(i)}_{max}\approx2\vert\alpha\vert^2\left(n+1\right). 
\end{equation}

\begin{figure}
	\includegraphics[scale=0.5]{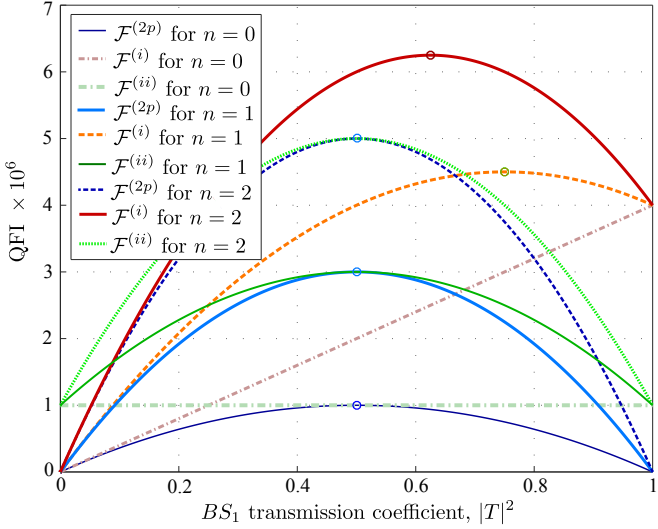}
	\caption{Coherent plus Fock state input for $n=0,1,2$ and $\vert\alpha\vert=10^3$. The Fock states can be seen as a ``boost'' to the coherent source in terms of QFI. Each circle marks the maximum of the corresponding curve. }
	\label{fig:Coh1_Fock0_alpha1000_n_012_versus_T}
\end{figure}
The symmetric single-parameter QFI \eqref{eq:F_ii_shorthand} is found to be
\begin{equation}
\label{eq:F_ii_Fock_plus_coherent}
\mathcal{F}^{(ii)}=
|\alpha|^2
+4n\left(1+2\vert\alpha\vert^2\right)\vert{TR}\vert^2
\end{equation}
and it is obviously maximal in the balanced case when we have $\mathcal{F}^{(ii)}_{max}=\mathcal{F}^{(2p)}_{max}=\vert\alpha\vert^2+n(1+2\vert\alpha\vert^2)$, result\footnote{In reference \cite{Pezze2013} the authors consider a slightly more complex input state, namely the separable density matrix input state (some notations have been adapted) $\hat{\rho}_{in}=\hat{\rho}_1\otimes\ket{n_0}\bra{n_0}$. They find the QFI $\mathcal{F}^{(ii)}=2\braket{\hat{n}_1}n+\braket{\hat{n}_1}+n$. For a coherent state at input port $1$ we have $\braket{\hat{n}_1}=|\alpha|^2$ implying the result from equation \eqref{eq:F_ii_Fock_plus_coherent} in the balanced case.} also found in reference \cite{Pezze2013}.

What is truly remarkable about the input state \eqref{eq:psi_in_Fock_plus_coherent} is the total absence of an input PMCs, regardless on the QFI considered. Once again, we can argue that this is due to the Fock states having no well defined phase \cite{GerryKnight}.

In Fig.~\ref{fig:Coh1_Fock0_alpha1000_n_012_versus_T} we depict the effect of adding $n=1$ and $n=2$ photons at input port $0$, while having a coherent state at input port $1$. We voluntarily used the high-intensity regime ($\vert\alpha\vert^2\gg1$) in order to show that the effect of a single photon in port $0$ is having a significant impact on all three QFIs in this regime. Adding more photons in port $0$ enhances the effect, while also displacing $T^{(i)}_{opt}$ towards the balanced case. This result could have been also anticipated from equation \eqref{eq:T_i_opt_C2prime_C4prime_ZERO_Fock_plus_coherent}, since for $\vert\alpha\vert^2\gg1$,
\begin{equation}
T^{(i)}_{opt}\approx\sqrt{\frac{1}{2}+\frac{1}{4n}}.
\end{equation}

\begin{figure}
	\includegraphics[scale=0.5]{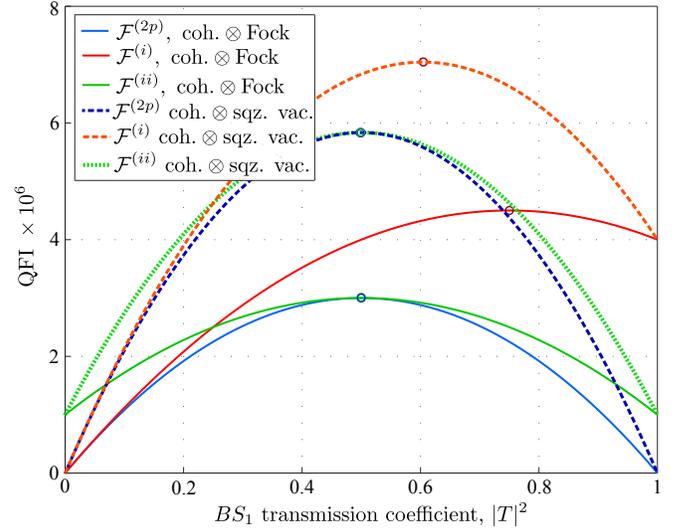}
	\caption{The coherent plus Fock input state versus the coherent plus squeezed vacuum input state. Parameters used: $\vert\alpha\vert=10^3$,  $n=1$ for the Fock input and $r=0.88$ ($\sinh^2r\approx1$) with the PMC $2\theta_\alpha-\theta=0$ for the squeezed vacuum input. Each circle marks the maximum of the corresponding curve.}
	\label{fig:CohFock_vs_CohSqzvac_alpha1k_n1_vs_T}
\end{figure}

While the results from Fig.~\ref{fig:Coh1_Fock0_alpha1000_n_012_versus_T} might seem impressive, it would be more useful to compare them with \emph{e. g.} the coherent plus squeezed vacuum input state \eqref{eq:psi_in_coh_sqzvac} under the constraint of having the same average number of input photons, \emph{i. e.} $n=\sinh^2{r}$. 

We thus keep the same coherent amplitude from Fig.~\ref{fig:Coh1_Fock0_alpha1000_n_012_versus_T} and choose $n=1$. In order to have a fair comparison we choose for the squeezed vacuum $r\approx0.88$, thus $\braket{n_0}=1.02\approx1$.

In Fig.~\ref{fig:CohFock_vs_CohSqzvac_alpha1k_n1_vs_T} we plot both scenarios for the same average number of input photons $\bar{N}=10^6+1\approx10^6$. The coherent plus squeezed vacuum input state outperforms the coherent plus Fock input for all three considered QFIs, with a factor of roughly $2$. This tendency continues and for $n\gg1$ we can easily estimate the ratio of the maximum QFIs \emph{e. g.} ${{\mathcal{F}_{max}^{(i)}}\vert_{coh-sqz}}/{{\mathcal{F}_{max}^{(i)}}\vert_{coh-Fock}}$. The optimal QFI $\mathcal{F}^{(i)}$ for a coherent plus squeezed vacuum input state in the high-coherent regime can be approximated as \cite{Ata20}
\begin{equation}
\mathcal{F}_{max}^{(i)}\vert_{coh-sqz}\approx{\vert\alpha\vert^2e^{2r}}
\approx4{\vert\alpha\vert^2\sqrt{n(n+1)}},
\end{equation}
while from equation \eqref{eq:F_i_Fock_plus_coherent_MAX} we have ${\mathcal{F}_{max}^{(i)}}\vert_{coh-Fock}\approx2n\vert\alpha\vert^2$. We end up with the ratio of the maximum QFIs,
\begin{equation}
\frac{{\mathcal{F}_{max}^{(i)}}\vert_{coh-sqz}}{{\mathcal{F}_{max}^{(i)}}\vert_{coh-Fock}}
\approx2\sqrt{\frac{n}{n+1}}.
\end{equation}
%
In conclusion, both input states display a similar scaling with the average number of input photons, with  the coherent plus squeezed vacuum input showing a better scaling coefficient. However, this advantage comes only at a cost, namely by precisely obeying the input PMC condition $2\theta_\alpha-\theta=0$. 

\subsection{Two-mode squeezed vacuum}
\label{subsec:TMSV}
As a last example we consider the two-mode squeezed vacuum (TMSV) \cite{GerryKnight,Anisimov2010} input state,
\begin{equation}
\label{eq:psi_in_TMSV}
\ket{\psi_{in}}
=\sum_{n=0}^\infty{\frac{(-1)^n}{\cosh{r}}(e^{i\theta}\tanh{r})^n\ket{n_1n_0}},
\end{equation}
also called the twin-beam state. This state can be elegantly obtained as $\ket{\psi_{in}}={\hat{S}_{tm}\left(\xi\right)}\ket{0_10_0}$, where the operator $\hat{S}_{tm}\left(\xi\right)$ defined by \cite{GerryKnight}
\begin{equation}
\hat{S}_{tm}\left(\xi\right)=e^{\xi^*{\hat{a}_0}{\hat{a}_1}-\xi{\hat{a}_0^\dagger}{\hat{a}_1^\dagger}}
\end{equation}
and can be seen as a two-mode analog of the squeezing operator \eqref{eq:Squeezing_operator}. Calculations are detailed in Appendix \ref{sec:app:TMSV_calculations} and for the two-parameter QFI we find
\begin{equation}
\label{eq:F_2p_two_mode_sqz_vac}
\mathcal{F}^{(2p)}=16\vert{TR}\vert^2\sinh^2r(1+\sinh^2r)
\end{equation}
thus $T^{(2p)}_{opt}=1/\sqrt{2}$ and $\mathcal{F}^{(2p)}_{max}=4\sinh^2r(1+\sinh^2r)$. The asymmetric single-parameter QFI $\mathcal{F}^{(i)}$ is found to be
\begin{equation}
\label{eq:F_i_two_mode_sqz_vac}
\mathcal{F}^{(i)}=\sinh^22r+16\vert{TR}\vert^2\sinh^2r(1+\sinh^2r)
\end{equation}
and the optimum transmission coefficient $T^{(i)}_{opt}$ is found in the balanced case, too, yielding $\mathcal{F}_{max}^{(i)}=8\sinh^2r(1+\sinh^2r)$. Finally, the symmetric single-parameter QFI $\mathcal{F}^{(ii)}$ is
\begin{equation}
\label{eq:F_ii_two_mode_sqz_vac}
\mathcal{F}^{(ii)}=16\vert{TR}\vert^2\sinh^2r(1+\sinh^2r)
\end{equation}
and $T^{(ii)}_{opt}=1/\sqrt{2}$. We immediately remark that in the case of $\mathcal{F}^{(ii)}$ there is no metrological advantage in having access to an external phase reference. At a second glance this should come as no surprise since the constraints \eqref{eq:No_metrological_advantage_Fii} are satisfied.

We can compare our results with previously reported ones. The total average number of input photons for TMSV is $\bar{N}=2\sinh^2r$, thus our previous results read
\begin{equation}
\label{eq:F_2p_Fii_two_mode_sqz_vac_N_bar}
\mathcal{F}^{(2p)}=\mathcal{F}^{(ii)}=4\vert{TR}\vert^2\bar{N}(\bar{N}+2)
\end{equation}
The optimum QFI is found in the balanced case, yielding $\mathcal{F}^{(2p)}=\mathcal{F}^{(ii)}=\bar{N}(\bar{N}+2)$ and this is indeed the result reported in reference \cite{Anisimov2010}. In reference \cite{Sahota2015} the authors report $\mathcal{F}^{(ii)}=0$ (see Table I.) for TMSV and a balanced interferometer, however they seem to have confused the quantum states before and  after $BS_1$.

We compare now the TMSV  with a coherent plus squeezed vacuum input state \eqref{eq:psi_in_coh_sqzvac}. We impose the same total average photon number for both states $\bar{N}=2\sinh^2r$, we thus choose $\vert\alpha\vert=\sinh{r}$ with the PMC $2\theta_\alpha-\theta=0$, configuration that yields the optimum performance \cite{Pez08,Ata18} for the input state \eqref{eq:psi_in_coh_sqzvac}.

\begin{figure}
	\includegraphics[scale=0.5]{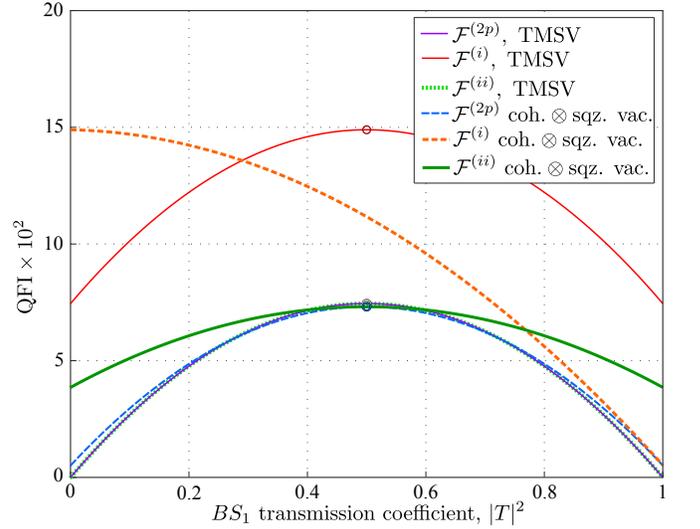}	
	\caption{The TMSV input state versus the coherent plus squeezed vacuum input state. Parameters used for TMSV: $r=2$ ($\sinh^2r\approx13$) and $\theta=0$. For the coherent plus squeezed vacuum input we used the same $r$, $\vert\alpha\vert=\sinh{r}$ and the PMC $2\theta_\alpha-\theta=0$. Each circle marks the maximum of the corresponding curve.}
	\label{fig:TMSV_vs_coh_sqzvac_Navg_26}
\end{figure}

In Fig.~\ref{fig:TMSV_vs_coh_sqzvac_Navg_26} we depict the performance of both input states. Since condition \eqref{eq:T_i_opt_is_0_COND_C2prime_C4prime_ZERO} is satisfied, for the coherent plus squeezed vacuum input state $T^{(i)}_{opt}=0$ and 
\begin{equation}
\mathcal{F}^{(i)}_{max}=4\Delta^2\hat{n}_0=2\bar{N}(\bar{N}+2)
\end{equation}
thus both input states yield the same maximum single-parameter QFI, $\mathcal{F}_{max}^{(i)}$, with the difference that while the TMSV reaches this value in the balanced case, the coherent plus squeezed vacuum attains it in the degenerate $T=0$ case. Another advantage of TMSV is its lower immunity wrt the value of $T$, its QFI $\mathcal{F}^{(i)}$ varying between  $\mathcal{F}^{(i)}_{max}/2$ and $\mathcal{F}^{(i)}_{max}$.

When it comes to the two-parameter QFI the performances of both input states are nearly identical, with an optimum transmission coefficient in the balanced case, $T^{(2p)}_{opt}=1/\sqrt{2}$.  Finally, for the symmetrical single-parameter QFI, $\mathcal{F}^{(ii)}$, the performances are nearly identical with an insignificant advantage for the coherent plus squeezed vacuum input state for values of $T$ far from the balanced case.

\section{Discussion}
\label{sec:discussion}

%


In this work we considered the QFI maximization for an unbalanced interferometer with a generic pure input state. However, since most reported works in the literature consider the balanced interferometric scenario, we briefly give now the conditions to have the maximum QFI in this case.

For the two-parameter QFI the necessary and sufficient condition for $T_{opt}^{(2p)}=1/\sqrt{2}$ is $C_2=0$ and $C_1>0$. For many input states this translates into well chosen input PMCs plus the condition that at least one of $\braket{\hat{a}_0}=0$ or $\braket{\hat{a}_1}=0$ must be satisfied. The same remarks apply to the symmetric single-parameter QFI, $\mathcal{F}^{(ii)}$. Thus, for the single-coherent, coherent plus squeezed vacuum, squeezed-coherent plus squeezed vacuum, twin-Fock, coherent plus Fock, TMSV -- to name just a few --, the maximum for both $\mathcal{F}^{(2p)}$ and $\mathcal{F}^{(ii)}$ is in the balanced scenario, with the appropriate input PMC (when applicable).

In the case of the asymmetric single-parameter QFI, $\mathcal{F}^{(i)}$, the conditions for having $T_{opt}^{(i)}=1/\sqrt{2}$ are more elaborate. For example, if $C'_2=C'_4=0$ and $C'_1>0$ (roughly equivalent to the previous conditions) one must also add $C'_3=0$ implying $\Delta^2\hat{n}_1=\Delta^2\hat{n}_0$. The same condition ($C'_3=0$) must be imposed to the input states yielding $C'_1=C'_2=0$. Thus, \emph{e. g.} for the twin Fock and TMSV the optimum QFI $\mathcal{F}^{(i)}_{max}$ is achieved for a balanced interferometer.

This paper addressed the maximization the QFI and this was done by choosing the appropriate transmission coefficient ($T$) for the first beam splitter. As discussed in Appendix \ref{sec:app:open_vs_closed_MZI}, the transmission coefficient of the second BS ($T'$) had no influence whatsoever on the QFI. However, this is not true when one considers a given detection scheme. Indeed, when optimizing the interferometric phase sensitivity for a specific detection scheme, $\Delta\varphi_{det}$, besides the input state, both $T$ and $T'$ come into play. For some input states, analytic formulas for $T'$ that optimize a specific input state and detection scheme can be found \cite{Ata20,Pre19}, but no general solution has been reported, to the best of our knowledge. We will address the optimization of $T'$ for a generic input state and for a number of given detection schemes in a future work.

\section{Conclusions}
\label{sec:conclusions}
In this paper we addressed the problem of finding the optimum transmission coefficient of the first beam splitter in a Mach-Zehnder interferometric setup in the sense of maximizing the quantum Fisher information. We addressed both the single- and two-parameter quantum Fisher information cases and gave closed-form analytical expressions for the optimum transmission coefficient for all discussed scenarios.

We also found the conditions needed to be fulfilled by the input state in order to render the availability of an external phase reference useless, irrespective of the value of the beam splitter transmission coefficient.

A number of input states were discussed and the $T$-dependence of each QFI assessed. Whenever possible, we compared our findings with the ones already reported in the literature. Among the considered input states, the squeezed-coherent plus squeezed-coherent input was shown to yield a significantly higher performance for the asymmetric single-parameter QFI, $\mathcal{F}^{(i)}$, than its two-parameter counterpart, $\mathcal{F}^{(2p)}$, if one uses the appropriate input phase matching conditions.

Possible evolutions of this work include the generalization to non-pure input states and taking into account losses.

%
%

\appendix

\section{The Fisher matrix coefficient $\mathcal{F}_{ss}$}
\label{sec:app:F_ss_calculations}
From equation \eqref{eq:F_ss_DEF} and using the fact that ${\vert\partial_{\varphi_s}\psi\rangle=-i\hat{G}_s\vert\psi\rangle}$, the sum-sum Fisher matrix element yields
\begin{equation}
\label{eq:F_ss_entangled}
\mathcal{F}_{ss}=\Delta^2{\hat{n}_0}+\Delta^2{\hat{n}_1}
+2\text{Cov}({\hat{n}_0},{\hat{n}_1})
\end{equation}
and, remarkably, $\mathcal{F}_{ss}$ is the only Fisher matrix coefficient not having a $T$-dependence. For a separable input state $\text{Cov}({\hat{n}_0},{\hat{n}_1})=0$ equation \eqref{eq:F_ss_entangled} thus simplifies to ${\mathcal{F}_{ss}=\Delta^2{\hat{n}_0}+\Delta^2{\hat{n}_1}}$. 

\begin{widetext}

\section{Shorthand notations}
\label{sec:app:Shorthand notations}
In order to improve readability, we introduce the following shorthand notations. For an entangled input state, we define:
\begin{equation}
\label{eq:VASP_shorthand_notations_entangled}
\left\{
\begin{array}{l}
V_{\pm}=\Delta^2{\hat{n}_0}\pm\Delta^2{\hat{n}_1}\\
V_{cov}=2\textrm{Cov}({\hat{n}_0},{\hat{n}_1})\\
A=4\Big(\langle{\hat{n}_0}\rangle+\langle{\hat{n}_1}\rangle
+2\big(\langle{\hat{n}_0}{\hat{n}_1}\rangle
-\vert\langle{\hat{a}_0^\dagger}{\hat{a}_1}\rangle\vert^2
-\Re\left\{\langle{(\hat{a}_0^\dagger)^2}{\hat{a}_1^2}\rangle
-\langle{\hat{a}_0^\dagger}{\hat{a}_1}\rangle^2\right\}
\big)\Big)\\
S_{\pm} = 4\Im\left\{\langle{\hat{a}_0^\dagger\hat{n}_0}{\hat{a}_1}\rangle
-\langle{\hat{n}_0}\rangle\langle{\hat{a}_0^\dagger}{\hat{a}_1}\rangle
\right\}
\pm4\Im\left\{
\langle{\hat{a}_0}{\hat{a}_1^\dagger\hat{n}_1}\rangle\}
-\langle{\hat{n}_1}\rangle\langle{\hat{a}_0}{\hat{a}_1^\dagger}\rangle
\right\}\\
P = 4\Im\left\{\langle{\hat{a}_0^\dagger}{\hat{a}_1}\rangle
\right\}
\end{array}
\right.
\end{equation}
where $\Im$ denotes the imaginary part. For a separable input state we have
\begin{equation}
\label{eq:VASP_shorthand_notations}
\left\{
\begin{array}{l}
V_\pm = \Delta^2{\hat{n}_0}\pm\Delta^2{\hat{n}_1}\\
A = 4\Big(\langle{\hat{n}_0}\rangle+\langle{\hat{n}_1}\rangle
+2\big(\langle{\hat{n}_0}\rangle\langle{\hat{n}_1}\rangle-\vert\langle{\hat{a}_0}\rangle\langle{\hat{a}_1}\rangle\vert^2
-\Re\left\{\langle{(\hat{a}_0^\dagger)^2}\rangle\langle{\hat{a}_1^2}\rangle
-\langle{\hat{a}_0^\dagger}\rangle^2\langle{\hat{a}_1}\rangle^2\right\}
\big)\Big)\\
S_\pm=4\Im\left\{\left(\langle{\hat{a}_0^\dagger\hat{n}_0}\rangle
-\langle{\hat{a}_0^\dagger}\rangle\langle{\hat{n}_0}\rangle\right)
\langle{\hat{a}_1}\rangle
\right\}
\pm4\Im\left\{\langle{\hat{a}_0}\rangle\left(\langle{\hat{a}_1^\dagger\hat{n}_1}\rangle
-\langle{\hat{n}_1}\rangle\langle{\hat{a}_1^\dagger}\rangle\right)
\right\}\\
P = 4\Im\left\{\langle{\hat{a}_0^\dagger}\rangle\langle{\hat{a}_1}\rangle
\right\}.
\end{array}
\right.
\end{equation}

\section{The Fisher matrix coefficient $\mathcal{F}_{dd}$}
\label{sec:app:F_dd_calculations}
From equation \eqref{eq:Fisher_matrix_elements} and using the fact that $\vert\partial_{\varphi_d}\psi\rangle=-i\hat{G}_d\vert\psi\rangle=-i\hat{J}_z\vert\psi\rangle$ we have $\mathcal{F}_{dd}=4\Delta^2\hat{J}_z$. We employ now the relation
\begin{equation}
\label{eq:exp_Jx_Jz_exp_minus_Jx}
e^{-i\vartheta{\hat{J}_x}}{\hat{J}_z}e^{i\vartheta{\hat{J}_x}}
=\cos\vartheta{\hat{J}_z}-\sin\vartheta{\hat{J}_y}
\end{equation}
to equations \eqref{eq:partial_varphi_d_psi_partial_varphi_d_psi_simplif} and \eqref{eq:psi_partial_varphi_d_psi_simplif}. After some straightforward algebra we are lead to
the result from equation
\eqref{eq:Fdd_cos2_vartheta_Variance_Jz_sin2_vartheta_Variance_Jy}. We also find by direct calculation
\begin{equation}
\label{eq:Variance_Jz_Jy_SymCov_JyJz}
\left\{
\begin{array}{l}
\Delta^2{\hat{J}_z}=\frac{1}{4}
\left(
\Delta^2{\hat{n}_0}+\Delta^2{\hat{n}_1}-2\text{Cov}\left({\hat{n}_0},{\hat{n}_1}\right)
\right)\\
\Delta^2{\hat{J}_y}=\frac{1}{4}
\Big(
2\left(\braket{{\hat{n}_0}{\hat{n}_1}}-\braket{{\hat{a}_0^\dagger}{\hat{a}_1}}\braket{{\hat{a}_0}{\hat{a}_1^\dagger}}\right)
+\braket{{\hat{n}_0}}
+\braket{{\hat{n}_1}}-2\Re\left\{\braket{{(\hat{a}_0^\dagger)^2}{\hat{a}_1^2}}
-\braket{{\hat{a}_0^\dagger}{\hat{a}_1}}^2\right\}
\Big)\\
\widehat{\text{Cov}}({\hat{J}_z},{\hat{J}_y})
=\frac{1}{2}\Im\left\{
\braket{{\hat{a}_0^\dagger\hat{n}_0}{\hat{a}_1}}
-\braket{{\hat{n}_0}}\braket{{\hat{a}_0^\dagger}{\hat{a}_1}}\right\}
+\frac{1}{2}\Im\left\{\braket{{\hat{a}_0}{\hat{a}_1^\dagger\hat{n}_1}}
-\braket{{\hat{n}_1}}\braket{{\hat{a}_0}{\hat{a}_1^\dagger}}
\right\}
\end{array}
\right.
\end{equation}
and since we parametrized $T=|T|=\cos\frac{\vartheta}{2}$ and $R=i\sqrt{1-|T|^2}$, we have $\cos^2\vartheta=\left(|T|^2-|R|^2\right)^2$, $\sin^2\vartheta=4|TR|^2$ and $\sin2\vartheta=2|TR|\left(|T|^2-|R|^2\right)$. Thus, the difference-difference Fisher matrix coefficient $\mathcal{F}_{dd}$ for an unbalanced interferometer is found to be
\begin{eqnarray}
\label{eq:F_dd_non_bal_entangled}
\mathcal{F}_{dd}
=\left(\vert{T}\vert^2-\vert{R}\vert^2\right)^2
\left(\Delta^2{\hat{n}_0}+\Delta^2{\hat{n}_1}
-2\text{Cov}\left({\hat{n}_0},{\hat{n}_1}\right)
\right)
\nonumber\\
+4\vert{TR}\vert^2
\left(
\langle{\hat{n}_1}\rangle
+\langle{\hat{n}_0}\rangle
+2\left(\langle{\hat{n}_0}{\hat{n}_1}\rangle
-|\langle{\hat{a}_0^\dagger}{\hat{a}_1}\rangle|^2
-\Re\{\langle{(\hat{a}_0^\dagger)^2}{\hat{a}_1^2}\rangle-\langle{\hat{a}_0^\dagger}{\hat{a}_1}\rangle^2\}
\right)
\right)
\nonumber\\
-8|TR|\left(\vert{T}\vert^2-\vert{R}\vert^2\right)\left(
\Im\left\{\langle{\hat{a}_0^\dagger\hat{n}_0}{\hat{a}_1}\rangle
-\langle{\hat{a}_0^\dagger}{\hat{a}_1}\rangle
\langle{\hat{n}_0}\rangle\right\}
+\Im\left\{\langle{\hat{a}_0}{\hat{a}_1^\dagger\hat{n}_1}\rangle\}
-\langle{\hat{a}_0}{\hat{a}_1^\dagger}\rangle
\langle{\hat{n}_1}\rangle\right\}
\right)
\end{eqnarray}
If we assume a separable input state, equation \eqref{eq:F_dd_non_bal_entangled} simplifies to \cite{Ata20},
\begin{eqnarray}
\label{eq:F_dd_FINAL_FORM_GENERAL_compact2}
\mathcal{F}_{dd}=\left(\vert{T}\vert^2-\vert{R}\vert^2\right)^2
\left(\Delta^2{\hat{n}_0}
+\Delta^2{\hat{n}_1}
\right)
\nonumber\\
+4\vert{TR}\vert^2
\left(\langle{\hat{n}_0}\rangle
+\langle{\hat{n}_1}\rangle
+2\left(\langle\hat{n}_0\rangle\langle\hat{n}_1\rangle
-\vert\langle{\hat{a}_0}\rangle\langle{\hat{a}_1}\rangle\vert^2
-\Re\left\{
\langle{(\hat{a}_0^\dagger)^2}\rangle\langle{\hat{a}_1^2}\rangle
-\langle{\hat{a}_0^\dagger}\rangle^2\langle{\hat{a}_1}\rangle^2
\right\}
\right)
\right)
\nonumber\\
-8|TR|\left(\vert{T}\vert^2-\vert{R}\vert^2\right)\left(
\Im\left\{\left(\langle{\hat{a}_0^\dagger\hat{n}_0}\rangle
-\langle{\hat{a}_0^\dagger}\rangle\langle{\hat{n}_0}\rangle\right)
\langle{\hat{a}_1}\rangle
+\langle{\hat{a}_0}\rangle\left(\langle{\hat{a}_1^\dagger\hat{n}_1}\rangle
-\langle{\hat{n}_1}\rangle\langle{\hat{a}_1^\dagger}\rangle\right)
\right\}
\right)
\end{eqnarray}
and wrt the cited reference we grouped some terms in order to make the $T$-dependence more obvious. Regardless if the input state is separable or entangled, the $\mathcal{F}_{dd}$ Fisher matrix element can be written in shorthand notations \eqref{eq:F_2p_shorthand} by using the identity $\left(\vert{T}\vert^2-\vert{R}\vert^2\right)^2=1-4|TR|^2$ where in the case of an entangled input state the notations from equation \eqref{eq:VASP_shorthand_notations_entangled} are employed, while in the case of a separable input state, the ones from \eqref{eq:VASP_shorthand_notations} should be used. If, moreover, we assume a balanced interferometer, $\mathcal{F}_{dd}$ reduces to
\begin{equation}
\mathcal{F}_{dd}=A=\langle{\hat{n}_0}\rangle
+\langle{\hat{n}_1}\rangle
+2\left(\langle{\hat{n}_0}\rangle\langle{\hat{n}_1}\rangle
-\vert\langle{\hat{a}_0}\rangle\langle{\hat{a}_1}\rangle\vert^2
-\Re\left\{
\langle{(\hat{a}_0^\dagger)^2}\rangle\langle{\hat{a}_1^2}\rangle
-\langle{\hat{a}_0^\dagger}\rangle^2\langle{\hat{a}_1}\rangle^2
\right\}
\right)
\end{equation}
and this expression is found in the literature \cite{Lan14,Ata19}, sometimes with supplimentary simplifying assumptions (\emph{e. g.} $\braket{\hat{a}_0}=0$ yielding equation (3) from reference \cite{Liu13} or equation (13) from reference \cite{Lan13} when assuming a coherent input state in port $1$).

\section{The Fisher matrix coefficient $\mathcal{F}_{sd}$}
\label{sec:app:F_sd_calculations}
Employing relation \eqref{eq:exp_Jx_Jz_exp_minus_Jx}, after some straightforward algebra we are led to the result from equation \eqref{eq:F_sd_Schwinger_FINAL}. Please note that we used the covariance \eqref{eq:covariance_DEF} and not the symmetrized covariance \eqref{eq:SymmetrizedCovariance_DEF} because $\hat{N}$ commutes with both $\hat{J}_y$ and $\hat{J}_z$. By direct calculation one finds
\begin{equation}
\label{eq:Cov_N_Jz_Cov_N_Jy}
\left\{
\begin{array}{l}
\text{Cov}\left({\hat{N}},{\hat{J}_z}\right)=\frac{1}{2}\left(\Delta^2{\hat{n}_0}-\Delta^2\hat{n}_1\right)\\
\text{Cov}\left(\hat{N},\hat{J}_y\right)
=\Im\left\{\braket{{\hat{a}_0^\dagger}{\hat{n}_0}{\hat{a}_1}}-\braket{{\hat{n}_0}}\braket{{\hat{a}_0^\dagger}{\hat{a}_1}}\right\}
-\Im\left\{\braket{{\hat{a}_0}{\hat{a}_1^\dagger}{\hat{n}_1}}-\braket{{\hat{a}_0}{\hat{a}_1^\dagger}}\braket{{\hat{n}_1}}\right\}
+\Im\left\{\braket{{\hat{a}_0^\dagger}{\hat{a}_1}}\right\}
\end{array}
\right.
\end{equation}
thus, wrt the input field operators the ``sum-difference'' Fisher matrix element is found to be
\begin{eqnarray}
\label{eq:F_sd_unbalanced_entangled_final}
\mathcal{F}_{sd}
=\left(\vert{T}\vert^2-\vert{R}\vert^2\right)
\left(\Delta^2\hat{n}_0
-\Delta^2\hat{n}_1
\right)
-4|TR|\Im\left\{\langle{\hat{a}_0^\dagger}{\hat{a}_1}\rangle\right\}
\nonumber\\
+4|TR|\left(\Im\left\{\langle{\hat{a}_0}{\hat{a}_1^\dagger\hat{n}_1}\rangle
-\langle{\hat{a}_0}{\hat{a}_1^\dagger}\rangle
\langle{\hat{n}_1}\rangle\right\}
-\Im\left\{\langle{\hat{a}_0^\dagger\hat{n}_0}{\hat{a}_1}\rangle
-\langle{\hat{a}_0^\dagger}{\hat{a}_1}\rangle\langle{\hat{n}_0}\rangle
\right\}
\right).
\end{eqnarray}
For a separable input state we get have the result \cite{Ata20},
\begin{eqnarray}
\label{eq:F_sd_GENERIC_FINAL}
\mathcal{F}_{sd}
=\left(\vert{T}\vert^2-\vert{R}\vert^2\right)
\left(\Delta^2\langle{\hat{n}_0}\rangle
-\Delta^2\langle{\hat{n}_1}\rangle
\right)
-4\vert{TR}\vert\Im\left\{\langle{\hat{a}_0^\dagger}\rangle\langle{\hat{a}_1}\rangle\right\}
\nonumber\\
-4\vert{TR}\vert\Im\left\{\left(\langle{\hat{a}_0^\dagger\hat{n}_0}\rangle
-\langle{\hat{a}_0^\dagger}\rangle\langle{\hat{n}_0}\rangle\right)\langle{\hat{a}_1}\rangle\right\}
+4\vert{TR}\vert\Im\left\{\langle{\hat{a}_0}\rangle
\left(\langle{\hat{a}_1^\dagger\hat{n}_1}\rangle
-\langle{\hat{a}_1^\dagger}\rangle
\langle{\hat{n}_1}\rangle\right)\right\}.
\end{eqnarray}

\section{The two-parameter QFI and the Fisher matrix}
\label{sec:app:QFI_Fisher_matrix}
Since we have a two-parameter estimation problem ($\varphi_d$ and $\varphi_s$) we are compelled to use the Fisher matrix \cite{Jar12}. Definition \eqref{eq:Fisher_matrix_elements} allows one to construct the $2\times2$ Fisher information matrix \cite{Paris2009,Lan13,Lan14},
\begin{equation}
\label{eq:Fisher_matrix}
   \mathcal{F}=
  \left[ {\begin{array}{cc}
  \mathcal{F}_{ss} & \mathcal{F}_{sd} \\
   \mathcal{F}_{ds} & \mathcal{F}_{dd} \\
  \end{array} } \right]
\end{equation}
and the quantum Cram\'er-Rao bound inequality implies \cite{Lan13,Pez15}
\begin{equation}
\label{eq:Sigma_matrix_greater_inv_Fisher_matrix}
  \left[ {\begin{array}{cc}
  \Delta^2\varphi_s & \text{Cov}(\varphi_s,\varphi_d) \\
   \text{Cov}(\varphi_s,\varphi_d) & \Delta^2\varphi_d \\
  \end{array} } \right]
  =\boldsymbol{\Sigma}\geq\boldsymbol{\mathcal{F}}^{-1}=\frac{1}{\mathcal{F}_{ss}\mathcal{F}_{dd}-\mathcal{F}_{sd}\mathcal{F}_{ds}}
  \left[ {\begin{array}{cc}
  \mathcal{F}_{dd} & -\mathcal{F}_{sd} \\
   -\mathcal{F}_{ds} & \mathcal{F}_{ss} \\
  \end{array} } \right].
\end{equation}
Generally, this matrix inequality \emph{i. e.} $\boldsymbol{\Sigma}\geq\boldsymbol{\mathcal{F}}^{-1}$ cannot be saturated for all components. However, we are solely interested in the difference-difference phase estimator, $\Delta\varphi_d$, thus the only inequality we are interested to saturate is
\begin{equation}
\label{eq:Delta_varphi_geq_Fisher_2p_dd}
\Delta^2\varphi_d\geq{(\boldsymbol{\mathcal{F}}^{-1})_{dd}}={\frac{\mathcal{F}_{ss}}{\mathcal{F}_{ss}\mathcal{F}_{dd}-\mathcal{F}_{sd}\mathcal{F}_{ds}}}
\end{equation}
and in order to simplify the writing we were led to introduce the definition from equation \eqref{eq:Fisher_information_F_2p_DEFINITION}.

\section{Open versus closed MZI}
\label{sec:app:open_vs_closed_MZI}
When it comes to estimating the QFI in a Mach-Zehnder interferometric setup, most authors simply disregard the second beam splitter \cite{Jar12,Lan13,Lan14,Takeoka2017} and consider the quantum state $\ket{\psi}$ (see Fig.~\ref{fig:MZI_Fisher_two_phases}) when applying the QFI definition \eqref{eq:Fisher_matrix_elements}. Other authors, though, consider the full interferometer (see Fig.~\ref{fig:MZI_2D_closed_measure_at_output}), some in the case of the classical Fisher information \cite{Pez08} (see also the supplementary material of \cite{Lan13}), but mostly in the case of QFI \cite{Liu13,Pez15,Yu2018}. Indeed, in the balanced case, starting from equation \eqref{eq:psi_out_U_BS_dagger_U_varphi_U_BS_psi_in_BALANCED} and due to the exponential form of the generator (\emph{i. e.} $\hat{U}_\varphi=e^{i\varphi\hat{G}}$, see reference \cite{Paris2009}) the QFI is simply \cite{Liu13,Pez15,Yu2018,Pezze2013,Hou2019}
\begin{equation}
\label{eq:QFI_is_4_times_Delta_J_y}
\mathcal{F}
=4\Delta^2{\hat{J}_y}
=4\left(\braket{\psi_{out}|\hat{J}_y^2|\psi_{out}}-\braket{\psi_{out}|\hat{J}_y|\psi_{out}}^2\right).
\end{equation}
In the following, we will show that when it comes to estimating the QFI, ignoring the second BS is justified. This assertion remains true even in the non-balanced case, with beam splitters having different transmission coefficients (\emph{i. e.} $\vartheta'\neq\vartheta$).  This remark is not true for the classical Fisher information, since one starts from the output conditional probabilities \cite{Paris2009}.

We focus on the difference-difference Fisher matrix element (see Section \ref{subsec:the_Fisher_matrix}), but all other evaluations pursue an identical route. From definition \eqref{eq:Fisher_matrix_elements} we have
\begin{equation}
\label{eq:Fisher_matrix_elem_Fdd}
\mathcal{F}_{dd}=4\left(\braket{\partial_{\varphi_d}\psi_{out}|\partial_{\varphi_d}\psi_{out}}
-|\braket{\psi_{out}|\partial_{\varphi_d}\psi_{out}}|^2\right).
\end{equation}
Evaluating $\ket{\partial_{\varphi_d}\psi_{out}}=\partial\ket{\psi_{out}}/\partial\varphi_d$ and considering the first term from equation \eqref{eq:Fisher_matrix_elem_Fdd} takes us to
\begin{eqnarray}
\label{eq:partial_varphi_d_psi_partial_varphi_d_psi}
\braket{\partial_{\varphi_d}\psi_{out}|\partial_{\varphi_d}\psi_{out}}
=\bra{\psi_{in}}\hat{U}^\dagger_{BS}\left(\vartheta\right)\hat{U}^\dagger_\varphi\hat{J}_z\hat{U}_{BS}\left(\vartheta'\right)
\hat{U}^\dagger_{BS}\left(\vartheta'\right)\hat{J}_z\hat{U}_\varphi\hat{U}_{BS}\left(\vartheta\right)\ket{\psi_{in}}
\end{eqnarray}
and unitarity implies $\hat{U}^\dagger_{BS}\left(\vartheta'\right)\hat{U}_{BS}\left(\vartheta'\right)=\mathbb{I}$ thus equation \eqref{eq:partial_varphi_d_psi_partial_varphi_d_psi} simplifies to
\begin{equation}
\label{eq:partial_varphi_d_psi_partial_varphi_d_psi_simplif}
\braket{\partial_{\varphi_d}\psi_{out}|\partial_{\varphi_d}\psi_{out}}
=\bra{\psi_{in}}\hat{U}^\dagger_{BS}\left(\vartheta\right)\hat{J}^2_z
\hat{U}_{BS}\left(\vartheta\right)\ket{\psi_{in}}.
\end{equation}
In this last expression we used the fact that $\hat{U}_\varphi$ commutes with both $\hat{J}_z$ and $\hat{N}$. A similar simplification applies to the second term of $\mathcal{F}_{dd}$,
\begin{equation}
\label{eq:psi_partial_varphi_d_psi_simplif}
\braket{\psi_{out}|\partial_{\varphi_d}\psi_{out}}
=\bra{\psi_{in}}\hat{U}^\dagger_{BS}\left(\vartheta\right)\hat{J}_z
\hat{U}_{BS}\left(\vartheta\right)\ket{\psi_{in}}
\end{equation}
and the operator $\hat{U}_{BS}\left(\vartheta'\right)$ modeling the second BS does not appear in the final expression of $\mathcal{F}_{dd}$. This remark equally applies to the partial derivatives in respect with $\varphi_s$. 
This is why starting from Section \ref{subsec:the_Fisher_matrix} we excluded $BS_2$ from our setup, arriving at the scheme usually found in the literature, namely Fig.~\ref{fig:MZI_Fisher_two_phases}.

\section{The single-parameter asymmetric QFI $\mathcal{F}^{(i)}$}
\label{sec:app:F_i_calculations}
From equation \eqref{eq:a2_a3_U_BS_dagger_a1_a0_U_BS} we obtain the field operator transformation
\begin{equation}
\label{eq:n3_versus_N_Jz_Jy}
\hat{n}_3
=\frac{\hat{N}}{2}-\cos\vartheta\hat{J}_z
+\sin\vartheta\hat{J}_y
\end{equation}
and by applying it to the definition \eqref{eq:F_i_is_four_times_variance_n3} we are led to
\begin{eqnarray}
\label{eq:Fi_n3_with_Schwinger}
\mathcal{F}^{(i)}
=\Delta^2\hat{N}
+4\cos^2\vartheta\Delta^2\hat{J}_z
+4\sin^2\vartheta\Delta^2\hat{J}_y
-4\sin2\vartheta\widehat{\text{Cov}}\left(\hat{J}_z,\hat{J}_y\right)
-4\cos\vartheta\text{Cov}\left(\hat{N},\hat{J}_z\right)
+4\sin\vartheta\text{Cov}\left(\hat{N},\hat{J}_y\right).
\end{eqnarray}
By comparing the above expression with equations \eqref{eq:F_ss_DEF}, \eqref{eq:Fdd_cos2_vartheta_Variance_Jz_sin2_vartheta_Variance_Jy} and \eqref{eq:F_sd_Schwinger_FINAL}, the relation \eqref{eq:F_i_is_Fss_plus_Fdd-2Fds} connecting $\mathcal{F}^{(i)}$ to the Fisher matrix coefficients is immediate. By replacing the variance/covariance terms via equations \eqref{eq:Variance_Jz_Jy_SymCov_JyJz} and \eqref{eq:Cov_N_Jz_Cov_N_Jy} we obtain
\begin{eqnarray}
\label{eq:F_i_nonbalanced_ENTANGLED_FINAL}
\mathcal{F}^{(i)}=
4\vert{R}\vert^4\Delta^2{\hat{n}_0}+4\vert{T}\vert^4\Delta^2{\hat{n}_1}
\nonumber\\
+4\vert{TR}\vert^2\left(
\langle{\hat{n}_0}\rangle
+\langle{\hat{n}_1}\rangle
+2\left(\text{Cov}({\hat{n}_0},{\hat{n}_1})
+(\langle{\hat{n}_0}{\hat{n}_1}\rangle
-|\langle{\hat{a}_0^\dagger}{\hat{a}_1}\rangle|^2)
-\Re\{\langle{(\hat{a}_0^\dagger)^2}{\hat{a}_1^2}\rangle
-\langle{\hat{a}_0^\dagger}{\hat{a}_1}\rangle^2\}
\right)
\right)
\nonumber\\
+16|TR|\vert{R}\vert^2\Im\left\{\left(\langle{\hat{a}_0^\dagger\hat{n}_0}{\hat{a}_1}\rangle
-\langle{\hat{n}_0}\rangle\langle{\hat{a}_0^\dagger}{\hat{a}_1}\rangle
\right)\right\}
-16|TR|\vert{T}\vert^2\Im\left\{
\left(
\langle{\hat{a}_0}{\hat{a}_1^\dagger\hat{n}_1}\rangle\}
-\langle{\hat{n}_1}\rangle\langle{\hat{a}_0}{\hat{a}_1^\dagger}\rangle
\right)
\right\}
+8|TR|\Im\{\langle{\hat{a}_0^\dagger}{\hat{a}_1}\rangle\}
\end{eqnarray}
If the input state is separable, we have the result \cite{Ata20},
\begin{eqnarray}
\label{eq:F_i_GENERAL}
\mathcal{F}^{(i)}
=4\vert{R}\vert^4\Delta^2\hat{n}_0
+4\vert{T}\vert^4\Delta^2\hat{n}_1
+4\vert{TR}\vert^2\left(\langle{\hat{n}_0}\rangle+\langle{\hat{n}_1}\rangle
+2\left(\langle{\hat{n}_0}\rangle\langle{\hat{n}_1}\rangle-\vert\langle{\hat{a}_0}\rangle\vert^2\vert\langle{\hat{a}_1}\rangle\vert^2
-\Re\left\{\langle{(\hat{a}_0^\dagger)^2}\rangle\langle{\hat{a}_1^2}\rangle
-\langle{\hat{a}_0^\dagger}\rangle^2\langle{\hat{a}_1}\rangle^2\right\}
\right)
\right)
\nonumber\\
+8\vert{TR}\vert\Im\left\{\langle{\hat{a}_0^\dagger}\rangle\langle{\hat{a}_1}\rangle\right\}
+16\vert{TR}\vert\vert{R}\vert^2\Im\left\{\left(\langle{\hat{a}_0^\dagger\hat{n}_0}\rangle
-\langle{\hat{a}_0^\dagger}\rangle\langle{\hat{n}_0}\rangle\right)\langle{\hat{a}_1}\rangle\right\}
-16\vert{TR}\vert\vert{T}\vert^2\Im\left\{\langle{\hat{a}_0}\rangle
\left(\langle{\hat{a}_1^\dagger\hat{n}_1}\rangle
-\langle{\hat{a}_1^\dagger}\rangle
\langle{\hat{n}_1}\rangle\right)\right\}
\qquad
\end{eqnarray}
We can use the identity $4\vert{T}\vert^2\Delta^2{\hat{n}_1}
+4\vert{R}\vert^2\Delta^2{\hat{n}_0}=2(\vert{T}\vert^2-\vert{R}\vert^2)(\Delta^2{\hat{n}_1}-\Delta^2{\hat{n}_0})+2 (\Delta^2{\hat{n}_1}+\Delta^2{\hat{n}_0})$ in order to write the above expressions in the form suitable for equation \eqref{eq:F_i_in_shorthand_C_prime}.

\section{The single-parameter asymmetric QFI $\mathcal{F}^{(i)}$ with the phase shift in the upper MZI arm}
\label{sec:app:F_i_calculations_n2}
In all our calculations from Section \ref{subsec:F_i} we considered our phase shift in the lower arm of our MZI, \emph{i. e.} $\varphi_1=0$ and $\varphi_2=\varphi$ in Fig.~\ref{fig:MZI_Fisher_two_phases}. Other authors might take the opposite setup with $\varphi_1=\varphi$ and $\varphi_2=0$. This implies the modification of the definition of the QFI to $\mathcal{F}^{(i)}_{(\hat{n}_2)}=4\Delta^2{\hat{n}_2}$. Using the field operator transformation ${\hat{n}_2} 
={\hat{N}}/{2}+\cos\vartheta{\hat{J}_z}-\sin\vartheta{\hat{J}_y}$ we are led to
\begin{eqnarray}
\label{eq:F_n2_Schwinger}
\mathcal{F}^{(i)}_{(\hat{n}_2)}
=\Delta^2{\hat{N}}
+4\cos^2\vartheta\Delta^2{\hat{J}_z}
+4\sin^2\vartheta\Delta^2{\hat{J}_y}
-4\sin2\vartheta\widehat{\text{Cov}}({\hat{J}_z},{\hat{J}_y})
+4\cos\vartheta\text{Cov}\left({\hat{N}},{\hat{J}_z}\right)
-4\sin\vartheta\text{Cov}\left({\hat{N}},{\hat{J}_y}\right)
\quad
\end{eqnarray}
and this time the relation connecting $\mathcal{F}^{(i)}_{(\hat{n}_2)}$ to the Fisher matrix elements is
\begin{equation}
\label{eq:F_i_n2_is_Fss_plus_Fdd_plus_2Fds}
\mathcal{F}^{(i)}_{(\hat{n}_2)}
=\mathcal{F}_{ss}+\mathcal{F}_{dd}+2\mathcal{F}_{sd}.
\end{equation}
The results in terms of maximal QFI remain unchanged, only the input PMCs have to be adapted.

\section{Calculations for the two-parameter QFI}
\label{sec:app:F_2p_calculations}
For an entangled input state, the Fisher matrix coefficient $\mathcal{F}_{dd}$ from equation \eqref{eq:F_dd_non_bal_entangled} can be put in the form
\begin{eqnarray}
\label{eq:F_dd_shorthand_notations_entangled}
\mathcal{F}_{dd}=
V_{+}-V_{cov}
+\vert{TR}\vert^2\left(A-4(V_{+}-V_{cov}\right)
+|TR|(|T|^2-|R|^2)S_{+}
\end{eqnarray}
and we employed the shorthand notations \eqref{eq:VASP_shorthand_notations_entangled}. If the input state is separable ($V_{cov}=0$) the above expression simplifies to
\begin{equation}
\label{eq:F_dd_shorthand_notations_separable}
\mathcal{F}_{dd}=
V_{+}+\vert{TR}\vert^2\left(A-4V_{+}\right)
+|TR|(|T|^2-|R|^2)S_{+}.
\end{equation}
%
%
\end{widetext}

For both entangled and separable input states $\mathcal{F}_{sd}$ from equation \eqref{eq:F_sd_shorthand_notations} can be put in the form
\begin{equation}
\label{eq:F_sd_shorthand_notations_abs_TR_only}
\mathcal{F}_{sd}
=(|T|^2-|R|^2)V_{-}
-\vert{TR}\vert\left(P
+S_{-}\right).
\end{equation}
Combining the appropriate $\mathcal{F}_{ss}$, $\mathcal{F}_{dd}$ and $\mathcal{F}_{sd}$ coefficients in shorthand notation allows us to write $\mathcal{F}^{(2p)}$ from equation \eqref{eq:F_2p_shorthand} where, for an entangled input state the coefficients are given by
\begin{equation}
\label{eq:F_2p_C_coefficients_entangled}
\left\{
\begin{array}{l}
C_0=V_{+}-V_{cov}
-\frac{V_{-}^2}{V_{+}+V_{cov}}\\
C_1={A}-4(V_{+}-V_{cov})+4\frac{V_{-}^2}{V_{+}+V_{cov}}
-\frac{\left(P+S_{-}\right)^2}{V_{+}+V_{cov}}\\
C_2=2\left(-S_{+}+\frac{\left(P+S_{-}\right)V_{-}}{V_{+}+V_{cov}}
\right)
\end{array}
\right.
\end{equation}
while for a separable input state they simplify to
\begin{equation}
\label{eq:F_2p_C_coefficients}
\left\{
\begin{array}{l}
C_0=4\frac{\Delta^2{\hat{n}_0}\Delta^2{\hat{n}_1}}{V_{+}}\\
C_1=A-16\frac{\Delta^2{\hat{n}_0}\Delta^2{\hat{n}_1}}{V_{+}}
-\frac{\left(P+S_{-}\right)^2}{V_{+}}\\
C_2=2\left(-S_{+}+\frac{\left(P
+S_{-}\right){V_{-}}
}{V_{+}}
\right).
\end{array}
\right.
\end{equation}
In order to find the optimum transmission coefficient, $T_{opt}$, we use equation \eqref{eq:T_and_R_as_fct_abs_TR} to write $\vert{T}\vert^2-\vert{R}\vert^2=\pm\sqrt{1-4\vert{TR}\vert^2}$ (``+'' if $\vert{T}\vert>\vert{R}\vert$), thus equation \eqref{eq:F_2p_shorthand} becomes
\begin{equation}
\label{eq:F_2p_TR_equation}
\mathcal{F}^{(2p)}=C_0+C_1\vert{TR}\vert^2\pm C_2\vert{TR}\vert\sqrt{1-4\vert{TR}\vert^2}.
\end{equation}
We seek the extrema of this function 
and find the solutions
\begin{equation}
\vert{TR}\vert_{sol}^2
=\frac{1}{8}\pm\frac{|C_1|}{8\sqrt{C_1^2+4C_2^2}}
\end{equation}
By solving the equation $|T|^2-|T|^4=\vert{TR}\vert_{sol}^2$ we have a double $\pm$ indeterminacy. Replacing the found solutions into equation \eqref{eq:F_2p_TR_equation} and using some simple arguments we eliminate the non-desired solutions ending up with the result from equation \eqref{eq:T_opt_squared_F_2p}.

\section{Calculations for the single-parameter QFI $\mathcal{F}^{(i)}$}
\label{sec:app:calc_T_i_opt_for_Fi}
Applying the notations from equation \eqref{eq:VASP_shorthand_notations_entangled} to the QFI \eqref{eq:F_i_nonbalanced_ENTANGLED_FINAL} yields the coefficients
\begin{equation}
\label{eq:F_i_C_prime_coeffs}
\left\{
\begin{array}{l}
C_0'=2V_{+}\\
C_1'=A-4(V_{+}-V_{cov})\\
C_2'=-2S_{+}\\
C_3'=-2{V_{-}}\\
C_4'=2(P+S_{-}).
\end{array}
\right.
\end{equation}
If the input state is separable, we employ the notations from \eqref{eq:VASP_shorthand_notations} to the QFI \eqref{eq:F_i_GENERAL} and the result is formally identical to the above one except that $V_{cov}=0$.

In order to find the optimum transmission coefficient in the most general case, we apply the replacement \eqref{eq:T_and_R_as_fct_abs_TR} to equation \eqref{eq:F_i_in_shorthand_C_prime} arriving at the result
\begin{eqnarray}
\mathcal{F}^{(i)}=C'_0+\vert{TR}\vert^2 C'_1
\mp|TR|\sqrt{1-4\vert{TR}\vert^2}C'_2
\nonumber\\
\mp\sqrt{1-4\vert{TR}\vert^2}C_3'+\vert{TR}\vert C_4'.
\end{eqnarray}
We consider now $\vert{TR}\vert$ as our variable and impose $\partial\mathcal{F}^{(i)}/\partial\vert{TR}\vert=0$. After some simple algebra we get the quartic equation
\begin{equation}
\label{eq:app:F_i_quartic_equations}
\mathcal{A}\chi^4
+\mathcal{B}\chi^3
+\mathcal{C}\chi^2
+\mathcal{D}\chi
+\mathcal{E}
=0
\end{equation}
where for readability we denote $\vert{TR}\vert=\chi$
and the coefficients are
\begin{equation}
\left\{
\begin{array}{l}
\mathcal{A}=16({C'_1}^2+4{C'_2}^2)\\
\mathcal{B}=16(4{C'_2}{C'_3}+{C'_1}{C'_4})\\
\mathcal{C}=4(4{C'_3}^2-4{C'_2}^2-{C'_1}^2+{C'_4}^2)\\
\mathcal{D}=-4(2{C'_2}{C'_3}+{C'_1}{C'_4})\\
\mathcal{E}={C'_2}^2-{C'_4}^2.
\end{array}
\right.
\end{equation}
Equation \eqref{eq:app:F_i_quartic_equations} is analytically solvable \cite{Abramowitz1972,Rees1922}. After finding the four solutions $\chi_{sol}$, it is likely that some results can be immediately removed by the conditions $\chi_{sol}\in\mathbb{R}$ and $\chi_{sol}\leq0.5$ (equivalent to $\vert{T}\vert\leq1$). For the remaining ones we have to solve $|TR|^2=\chi^2_{sol}$ and using the identity $|R|^2=1-|T|^2$ 
we immediately arrive at equation \eqref{eq:T_opt_for_F_i_general_case}.

\section{The two-parameter QFI for the single Fock input}
\label{sec:app:single_Fock_calculations}
One can argue that $\mathcal{F}^{(2p)}$ from equation \eqref{eq:F_2p_single_Fock_input} is meaningless because $\mathcal{F}_{ss}$ and $\mathcal{F}_{sd}$ for the input state  \eqref{eq:psi_in_single_Fock_input} are null, we are thus in a $0/0$ situation while applying definition \eqref{eq:Fisher_information_F_2p_DEFINITION}. We can avoid this inconvenience by assuming an input state slightly different from equation \eqref{eq:psi_in_single_Fock_input}, namely by applying a small coherent amplitude in port $0$,
\begin{equation}
\label{eq:psi_in_Fock1_Beta0}
\vert\psi_{in}\rangle=\vert n_1\beta_0\rangle.
\end{equation}
This is actually the scenario discussed in Section \ref{subsec:coh_plus_Fock_input}, however with the input ports inverted. We have the result $\mathcal{F}^{(2p)}=4|TR|\left(n+\vert\beta\vert^2+2n\vert\beta\vert^2\right)$ and by applying the limit $\beta\to0$ equation \eqref{eq:F_2p_single_Fock_input} is immediate.

\section{Calculations for the squeezed-coherent plus squeezed vacuum input}
\label{sec:app:sqzcoh_sqzvac_calculations}
Using the previously found results from equations \eqref{eq:Average_n1_sqz_coh_VACUUM0} and \eqref{eq:Variance_n1_sqz_coh_plus_sqz_vac_VACUUM0} we find the shorthand notations
\begin{equation}
\label{eq:app:VASP_shorthand_notations_sqzcoh_sqzvac}
\left\{
\begin{array}{l}
V_\pm=\frac{\sinh^22r}{2}\pm\frac{\sinh^22z}{2}
\nonumber\\
\qquad\pm{\vert\alpha\vert^2}\left(\cosh2z
-\sinh2z\cos\left(2\theta_\alpha-\phi\right)\right)\\
A=4\Big(\vert\alpha\vert^2\left(\cosh2r+\sinh2r\cos(2\theta_\alpha-\theta)\right)
\\
\qquad+\frac{\cosh2r\cosh2z+\sinh2r\sinh2z\cos(\theta-\phi)-1}{2}
\Big)\\
S_{\pm} = P = 0.
\end{array}
\right.
\end{equation}
Since $C_2=0$, the optimum for the two-parameter QFI occurs in a balanced interferometer. For the single-parameter QFI, we insert the above results into \eqref{eq:F_i_C_prime_coeffs} to get the $C'$-coefficients. If we impose the optimum input PMC \eqref{eq:PMC_sqz-coh_plus_sqz-vac} we find the optimum $BS_1$ transmission coefficient \cite{Ata20},
\begin{eqnarray}
\label{eq:T_i_opt_C2prime_C4prime_ZERO_sqzcoh_sqz_vac_PMC}
\left(T^{(i)}_{opt}\right)^2
=\frac{1}{2}
\nonumber\\
+\frac{1}{4}\frac{\sinh^22z-\sinh^22r
+2{\vert\alpha\vert^2}e^{2z}}
{\vert\alpha\vert^2\left(e^{2r}-e^{2z}
\right)
-\sinh^2(r+z)\cosh2(r-z)}.
\end{eqnarray}

\section{Calculations for the squeezed-coherent plus squeezed-coherent input}
\label{sec:app:sqzcoh_sqzcoh_calculations}
In order to compute the shorthand notations \eqref{eq:VASP_shorthand_notations} for a squeezed-coherent plus squeezed-coherent input state we first need to assess some terms appearing in these expressions. The variance for a squeezed-coherent state in input port 1 was already given in equation \eqref{eq:Variance_n1_sqz_coh_plus_sqz_vac_VACUUM0}. Similar calculations for a squeezed-coherent state in input port $0$ and be easily done \cite{Ata19} and combining these results gives
\begin{eqnarray}
\label{eq:app:shorthand_V_sqzcoh_sqzcoh}
V_\pm=\frac{\sinh^2{2r}}{2}+{\vert\beta\vert^2}\left(\cosh{2r}-\sinh{2r}\cos(2\theta_\beta-\theta)\right)
\nonumber\\
\pm\frac{\sinh^22z}{2}
\pm{\vert\alpha\vert^2}\left(\cosh2z
-\sinh2z\cos\left(2\theta_\alpha-\phi\right)\right).
\quad
\end{eqnarray}
For the second term, after some calculations we find
\begin{eqnarray}
\label{eq:app:shorthand_A_sqzcoh_sqzcoh}
A={\vert\beta\vert^2}\left(\cosh{2z}+\sinh{2z}\cos(2\theta_\beta-\phi)\right)
\nonumber\\
+{\vert\alpha\vert^2}\left(\cosh{2r}+\sinh{2r}\cos(2\theta_\alpha-\theta)\right)
\nonumber\\
+\frac{\cosh{2r}\cosh{2z}-\sinh{2r}\sinh{2z}\cos(\theta-\phi)-1}{2}.
\quad
\end{eqnarray}
We also have the results \cite{Ata19},
\begin{equation}
\left\{
\begin{array}{l}
\langle{\hat{n}_1\hat{a}_1}\rangle
-\langle{\hat{n}_1}\rangle\langle{\hat{a}_1}\rangle
=\alpha\sinh^2z-\frac{\alpha^*}{2}\sinh2ze^{i\phi}\\
\langle{\hat{n}_0\hat{a}_0}\rangle
-\langle{\hat{n}_0}\rangle\langle{\hat{a}_0}\rangle
=\beta\sinh^2r-\frac{\beta^*}{2}\sinh2re^{i\theta}
\end{array}
\right.
\end{equation}
thus
\begin{eqnarray}
\label{eq:app:shorthand_S1_sqzcoh_sqzcoh}
S_\pm
=2\vert\alpha\beta\vert\big(2\left(\sinh^2r\mp\sinh^2z\right)\sin(\theta_\alpha-\theta_\beta)
\nonumber\\
-\sinh2r\sin(\theta_\alpha+\theta_\beta-\theta)
\nonumber\\
\mp\sinh2z\sin(\theta_\alpha+\theta_\beta-\phi)
\big).
\end{eqnarray}
Finally, we find
\begin{equation}
\label{eq:app:shorthand_P_sqzcoh_sqzcoh}
P=4\vert\alpha\beta\vert\sin(\theta_\alpha-\theta_\beta).
\end{equation}
If we impose (PMC1) \emph{i. e.} equations \eqref{eq:PMC_sqz-coh_plus_sqz-vac} and \eqref{eq:theta_alpha_minus_theta_beta_ZERO}, the shorthand notations read
\begin{equation}
\label{eq:app:shorthand_VASP_sqzcoh_sqzcoh_PMC1}
\left\{
\begin{array}{l}
V_\pm=\frac{\sinh^2{2r}}{2}+{\vert\beta\vert^2}e^{-2r}\pm\frac{\sinh^22z}{2}\pm{\vert\alpha\vert^2}e^{2z}\\
A=4\left({\vert\beta\vert^2}e^{-2z}+{\vert\alpha\vert^2}e^{2r}+\sinh^2(r+z)\right)\\
S_\pm=P=0
\end{array}
\right.
\end{equation}
and we immediately have $C_2=C''_2=0$ implying that both $\mathcal{F}^{(2p)}$ and $\mathcal{F}^{(ii)}$ are optimized in the balanced case under the constraints $C_1>0$ and, respectively, $C''_1>0$. If we assume (PMC2) from equation \eqref{eq:PMC2} we end up with the coefficients
\begin{equation}
\label{eq:app:shorthand_VASP_sqzcoh_sqzcoh_PMC2}
\left\{
\begin{array}{l}
V_\pm=\frac{\sinh^2{2r}}{2}+{\vert\beta\vert^2}e^{-2r}\pm\frac{\sinh^22z}{2}\pm{\vert\alpha\vert^2}e^{-2z}\\
A=4\left({\vert\beta\vert^2}e^{2z}+{\vert\alpha\vert^2}e^{2r}+\sinh^2(r-z)\right)\\
S_\pm=P=0
\end{array}
\right.
\end{equation}
and again, since $C_2=C''_2=0$ both $\mathcal{F}^{(2p)}$ and $\mathcal{F}^{(ii)}$ are optimized in the balanced case if the constraints $C_1>0$ and, respectively, $C''_1>0$ are met.

Finally, if we assume (PMC3) we find the coefficients
\begin{equation}
\label{eq:app:shorthand_VASP_sqzcoh_sqzcoh_PMC3}
\left\{
\begin{array}{l}
V_\pm=\frac{\sinh^2{2r}}{2}+{\vert\beta\vert^2}e^{2r}\pm\frac{\sinh^22z}{2}\pm{\vert\alpha\vert^2}e^{2z}\\
A=4\left({\vert\beta\vert^2}e^{2z}+{\vert\alpha\vert^2}e^{2r}+\sinh^2(r+z)\right)\\
S_\pm=2\vert\alpha\beta\vert\left(2(\sinh^2r\mp\sinh^2z)+\sinh2r\pm\sinh2z\right)\\
P=4\vert\alpha\beta\vert
\end{array}
\right.
\end{equation}
and this time none of the QFIs is necessarily maximized in the balanced case.

\section{Calculations for the two-mode squeezed vacuum input}
\label{sec:app:TMSV_calculations}
We recall the fundamental relations needed to work with TMSV states \cite{GerryKnight},
\begin{equation}
\label{eq:TMSV_fundamental_relations}
\left\{
\begin{array}{l}
\hat{S}^\dagger_{tm}\left(\xi\right)\hat{a}_0\hat{S}_{tm}\left(\xi\right)
=\cosh{r}{\hat{a}_0}-\sinh{r}e^{i\theta}{\hat{a}_1^\dagger}\\
\hat{S}^\dagger_{tm}\left(\xi\right)\hat{a}_1\hat{S}_{tm}\left(\xi\right)
=\cosh{r}{\hat{a}_1}-\sinh{r}e^{-i\theta}{\hat{a}_0^\dagger}
\end{array}
\right.
\end{equation}
Since the input state is entangled we use now equations \eqref{eq:VASP_shorthand_notations_entangled} as definitions and get
\begin{equation}
\label{eq:VASP_shorthand_notations_two_mode_sqz_vac}
\left\{
\begin{array}{l}
V_+ = \frac{\sinh^22r}{2}\\
V_- = 0\\
V_{cov}=\frac{\sinh^22r}{2}\\
A=16\sinh^2r\cosh^2r
\\
S_+ = S_- = P = 0.
\end{array}
\right.
\end{equation}
Through straightforward calculations we find the averages
\begin{equation}
\label{eq:Average_n0_n1_two_mode_sqz_vac}
\braket{\hat{n}_0}
=\braket{\hat{n}_1}
=\sinh^2r
\end{equation}
and the variances
\begin{equation}
\label{eq:Variance_n0_n1_two_mode_sqz_vac}
\Delta^2{\hat{n}_0}
=\Delta^2{\hat{n}_1}
=\frac{\sinh^22r}{4}.
\end{equation}
Since this input state is entangled, we expect $\text{Cov}({\hat{n}_0},{\hat{n}_1})\neq0$. We find by direct calculation
\begin{equation}
\label{eq:Average_n1n0_two_mode_sqz_vac}
\braket{\hat{n}_0{\hat{n}_1}}
=(\cosh^2r+\sinh^2r)\sinh^2r
\end{equation}
and by employing equation \eqref{eq:Average_n0_n1_two_mode_sqz_vac} the covariance is found to be
\begin{equation}
\label{eq:Covariance_n1n0_two_mode_sqz_vac}
\text{Cov}({\hat{n}_0},{\hat{n}_1})
=\frac{\sinh^22r}{4}.
\end{equation}
For the two-parameter QFI we compute the $C$-coefficients from equation \eqref{eq:F_2p_C_coefficients_entangled} and have 
\begin{equation}
\label{eq:F_2p_C_coefficients_two_mode_sqz_vac}
\left\{
\begin{array}{l}
C_0=0\\
C_1=16\sinh^2r\cosh^2r
\\
C_2=0
\end{array}
\right.
\end{equation}
and we immediately get the result from equation \eqref{eq:F_2p_two_mode_sqz_vac}. Inserting the shorthand notations \eqref{eq:VASP_shorthand_notations_two_mode_sqz_vac} into equation \eqref{eq:F_i_C_prime_coeffs} takes us to the $C'$ coefficients
\begin{equation}
\label{eq:F_i_C_prime_coeffs_two_mode_sqz_vac}
\left\{
\begin{array}{l}
C_0'=\sinh^22r\\
C_1'=16\sinh^2r\cosh^2r\\
C_2'=C_3'=C_4'=0
\end{array}
\right.
\end{equation}
we thus find $\mathcal{F}^{(i)}$ from equation \eqref{eq:F_i_two_mode_sqz_vac}. Finally, from equation \eqref{eq:F_ii_C_sec_coeffs} we have
\begin{equation}
\label{eq:F_ii_C_sec_coeffs_two_mode_sqz_vac}
\left\{
\begin{array}{l}
C''_0=0\\
C''_1=16\sinh^2r\cosh^2r\\
C''_2=0
\end{array}
\right.
\end{equation}
yielding the symmetric single-parameter QFI from equation \eqref{eq:F_ii_two_mode_sqz_vac}.

\bibliography{MZI_phase_sensitivity_bibtex}

\end{document}